\documentclass[10pt,twocolumn]{revtex4-1}
\usepackage{amsmath,amssymb}
\usepackage{hyperref}
\usepackage{filecontents}
\usepackage{graphicx}
\usepackage{dcolumn}
\usepackage{bm}
\usepackage{subfigure}
\usepackage{graphicx}
\usepackage{braket}
\usepackage{xcolor}

\begin{document}


\title{The Gor'kov and Melik-Barkhudarov correction to an imbalanced Fermi gas in the presence of impurities}

\author{Heron Caldas} 
\email{hcaldas@ufsj.edu.br}
\affiliation{Departamento de Ci\^{e}ncias Naturais, Universidade Federal de S\~{a}o Jo\~{a}o Del Rei, Pra\c{c}a Dom Helv\'{e}cio 74, 36301-160, S\~{a}o Jo\~{a}o Del Rei, MG, Brazil}

\author{S. Rufo}
\email{sabrina.rufo11@gmail.com}
\author{M. A. R. Griffith}
\email{griffithphys@gmail.com}
 \affiliation{Beijing Computational Science Research Center, Beijing, 100193, China}
\affiliation{CeFEMA, Instituto Superior T\'ecnico, Universidade de Lisboa, Av. Rovisco Pais, 1049-001, Lisbon, Portugal}

\date{\today}



\begin{abstract} 
The effects of induced interactions are calculated in both clean and dirty situations, for balanced and imbalanced Fermi gases. We investigate the effects of nonmagnetic impurities on the induced interactions corrections to the transition temperature in the case of a balanced gas, and to the tricritical point in the case of an imbalanced Fermi gas at unitarity. We find that impurities act in detriment of the induced interactions, or particle-hole fluctuations, for the transition temperature and the tricritical point. For large impurity parameter, the particle-hole fluctuations are strongly suppressed. We have also found the Chandrasekhar-Clogston limit of an imbalanced Fermi gas at unitarity considering the effects of the induced interactions, both in the pure and impurity regimes.
\end{abstract}

\maketitle

\section{Introduction}

Superconductivity is a fascinating quantum phenomenon in which electrons form pairs and flow without resistance. High-temperature superconductors (HTS), materials that, at ambient pressure, behave as superconductors at temperatures above nearly $77 K$, the boiling point of liquid nitrogen, are expected to play a prominent role in emerging power technologies. However, the experimental processing of superconductive materials results in impurities in the material, which can affect its performance. Most of the detrimental effects of impurities are seen in the superfluid transition temperature $T_c$, which can be drastically decreased, when compared to the pure system~\cite{Fukuzumi}. While weak nonmagnetic impurities are not deleterious to $s$-wave BCS superconductors, according to Anderson's theorem,~\cite{Anderson,AndersonTheorem}, it has been observed that strong concentrations of nonmagnetic impurities, such as Zn, suppresses $T_c$~\cite{ReviewImpurities}. Consequently, the comprehension of how impurities affect $T_c$ is of fundamental importance to ultracold superfluids and high temperature superconductors~\cite{Review}. In this way, understanding the behavior of impurities in a given many-body system is a manner to actually probe and realize that system. It is then of crucial significance in condensed matter to know about the role of impurities on the normal to superconductor transition, and with cold atoms, experimental physicists currently have the ability to realize rather perfect Hamiltonians with which to test many-body physics~\cite{MartinThesis,MartinComment}. 

In addition to condensed matter, the comprehension of the effects of impurities has implications to other areas of physics such as nuclear physics, for example. In the core of neutron stars protons can be considered as impurities in the neutron matter, since the number density of protons is much less than that of neutrons~\cite{Wiringa}. It has been found that the {\it nuclear polaron problem} shares very interesting similarities to the one of polarons in solids, such as polaron localization~\cite{Kutschera}, and to that of Fermi polarons in an ultracold gas of fermionic atoms in the unitary limit, as an effective mass due to the dressing of the polaron by the medium~\cite{Kutschera,Forbes,Roggero,Nakano,Isaac}.

Besides impurity scattering, another important factor that has a strong impact on the superfluid phase is the Gorkov and Melik-Barkhudarov (GMB) screening of the interaction. The many-body effects, or simply the effects of the medium, on the two-body interaction, referred to as induced interactions, is responsible for the suppression of $T_c$ by a factor $(4e)^{1/3} \approx 2.2$ in a dilute balanced three dimensional (3D) spin-1/2 Fermi gas, compared with the (overestimated) mean-field (MF) BCS result, obtained without the GMB screening. This effect was considered first by Gor'kov and Melik-Barkhudarov~\cite{Gorkov}. However,  the factor $2.2$ is not an universal value for the GMB suppression. Dynamical mean-field theory (DMFT) calculations by Toschi et. al., obtained $T_c^{BCS}/T_c^{GMB} \simeq 3.13$~\cite{Toschi} in 2D lattices. In quasi 2D imbalanced Fermi gas it was found $T_c^{BCS}/T_c^{GMB} \simeq 2.72$~\cite{Resende}. In Ref.~\cite{Torma} at lower filling factors in a 2D lattice it was found $\Delta^{BCS}/\Delta^{GMB} \simeq 5$ and near half filling $\Delta^{BCS}/\Delta^{GMB} \simeq 10$. In 3D lattices, the authors of Ref.~\cite{Torma} have found that the screening effect of the induced interaction becomes stronger, $\Delta^{BCS}/\Delta^{GMB} \simeq 25$. These results show that the suppression of the order parameter is strongly enhanced by lattice effects (such as its geometry), as the filling factor becomes higher~\cite{Torma}, and also depends considerably on the coupling strength~\cite{Toschi}.

Indeed, it is well known that the consideration of the particle-hole channel has a significant effect in the solution of $T_\text{c}$ in the BCS--BEC crossover \cite{Review,Qijin}. This happens because there is a modification in the coupling of the interaction as a result of the screening of the interspecies interaction potential~\cite{Gorkov}. On the BEC side of the resonance (where there is a divergence of the two-body scattering length~\cite{note}) fluctuations in the pairing channel are predominant, while the GMB corrections become irrelevant deep in the BEC region and generally are taken as vanishing in this regime due to the destruction of the Fermi surface.

Usually, the computation of the GMB fluctuations has been constrained only to the balanced case i.e., with the ``Zeeman magnetic field'' $h \equiv (\mu_{\uparrow} - \mu_{\downarrow})/2$ set to zero, where $\mu_{\uparrow}$ and $\mu_{\downarrow}$ are the chemical potential of the spin-up and spin-down (non-interacting) fermions, respectively. Spin-imbalanced Fermi gases with a finite $h$ exhibit rich scenarios~\cite{Zwerger}. Starting with a superfluid (BCS) phase at zero temperature with $h = 0$, adding one of the spin species (spin-down atoms for instance) will increase $h$ and for $h=h_c$ there will be a first-order quantum phase transition between a pure BCS phase and a system with phase separated superfluid and normal states~\cite{PS1,PS2}. Experiments in 3D have observed phase separation between the superfluid and normal phases in the trapped gas~\cite{Exp1,Exp2,Exp3}.

Preceding studies found in the literature with $h \neq 0$ were concentrated on the effects of the induced interactions on the tricritical point $(p_\text{t},T_\text{t}/T_\text{F})$ of (pure) imbalanced Fermi gases in 3D~\cite{Yu10} and 2D~\cite{Resende}. Here $p_\text{t}=\left(n_{\uparrow}-n_{\downarrow}/n_{\uparrow}+n_{\downarrow}\right)_\text{t}$ and $T_\text{t}$ are the polarization and the temperature at the tricritical point, where $n_{\uparrow,\downarrow}$ are the spin densities and $T_\text{F}$ is the Fermi temperature. (We use the standard definitions of Fermi energy and Fermi momentum, given by $E_\text{F} =k_\text{B}T_\text{F} = \hbar^2 k_\text{F}^2 /2m$). See, for instance, Ref.~\cite{Qijin} for calculations taking into account both particle-hole and particle-particle contributions in clean systems.

As mentioned earlier, the consequences of nonmagnetic impurities in the gap parameter and critical temperature of s-wave superconductors are well known: weak impurities have no effects (due to Anderson's theorem), and under strong random nonmagnetic impurities (where Anderson's theorem is not applicable) the system breaks up into superconducting islands separated by an insulating sea~\cite{Nandini}. However, the effects of impurities on the medium (in which the particle-hole fluctuations give rise to the screening of the interactions), have not been addressed so far.

In this paper, we investigate the effects of the GMB corrections to mean-field results of Fermi gases with imbalanced spin populations, in the clean limit and in the presence of nonmagnetic impurities. We study the effects of impurities on the induced interactions corrections and its consequences to $T_c$, in the case of a balanced gas, and to the tricritical point in the case of an imbalance Fermi gas. The basic principle considered here is that the particle-hole fluctuations in the medium can themselves be affected by the presence of the impurity atoms.

The paper is organized as follows. In Section~\ref{M-H} the Hamiltonian and the thermodynamic potential of the model considered are presented. In Section~\ref{Ind} we calculate the induced interaction in the presence of impurities. The effects of the induced interactions in the presence of impurities on the critical temperature of a balanced Fermi gas are studied in Section~\ref{Tc}. In Section~\ref{TcP} we investigate the effects of the induced interactions in the presence of impurities on the tricritical point of an imbalanced Fermi gas. In Section~\ref{CCL} we find the Chandrasekhar-Clogston limit of an imbalanced Fermi gas at unitarity, considering the effects of the induced interactions in the pure and impurity regimes, by means of the T-matrix approximation. In Section~\ref{PE} we briefly comment on some current and possible experiments with impurities in cold atoms. We conclude in Section~\ref{Conc}.

\section{The Model Hamiltonian and Basic Definitions}
\label{M-H}

Let us begin with the (grand canonical) Hamiltonian of a 3D system of fermions interacting via an effective, short range pairing interaction $g$, with $g<0$, in momentum space
%
\begin{eqnarray}
H =  \sum_{\mathbf{k}\sigma} \xi_{\vec{k}\sigma}
c^{\dag}_{\mathbf{k}\sigma} c^{\ }_{\mathbf{k}\sigma} + g \sum_{\mathbf{k}\mathbf{k}'} 
c^{\dag}_{\mathbf{k} \uparrow} 
c^{\dag}_{-\mathbf{k} \downarrow} 
c^{\ }_{-\mathbf{k}' \downarrow} 
c^{\ }_{\mathbf{k}' \uparrow},
\label{Hamiltonian}
\end{eqnarray}
where $\xi_{\vec{k}\sigma} = e_k -\mu_\sigma = {k}^2/2m -\mu_\sigma$ refers to the bare dispersion relation of the spin-$\sigma = \uparrow, \downarrow$ fermions with mass $m$, $\mu_\sigma = \mu \pm h$ is the chemical potential of the species $\sigma$, and $\mu = (\mu_\uparrow + \mu_\downarrow)/2$ is the average of the $\uparrow$ and $\downarrow$ chemical potentials. Here $c^\dag$ ($c$) is the fermion creation (annihilation) operator. We have put the system volume to unity, and we also use the natural units, $\hbar = k_\text{B} = 1$.

At the mean-field (MF) level, the reduced Hamiltonian describing pairing between $\mathbf{k}$ and $-\mathbf{k}$ states is given by
\begin{eqnarray}
H^\text{MF}&=&\sum_{\mathbf{k}} \Big\{\xi_{\vec{k} \uparrow}
c_{\mathbf{k}\uparrow}^{\dag}c_{\mathbf{k}\uparrow} 
+ \xi_{\vec{k} \downarrow}
c_{-\mathbf{k} \downarrow}^{\dag}c_{-\mathbf{k} \downarrow} 
 \nonumber \\
&+&{}\Delta c_{-\mathbf{k} \downarrow}^{\dag}
 c_{\mathbf{k}\uparrow}^{\dag}+\Delta^*
 c_{\mathbf{k}\uparrow}c_{-\mathbf{k} \downarrow}\Big\} - \frac{|\Delta|^2}{g}\,.
\label{eq:19}
\end{eqnarray}
Here $\Delta = g\sum_\mathbf{k} \langle c_{\mathbf{k}\downarrow}c_{-\mathbf{k}\uparrow}\rangle$ is the MF order parameter. 

With $H^\text{MF}$ we obtain the thermodynamic potential $\mathcal{W}$, from which all thermodynamical relevant quantities can be obtained,

\begin{eqnarray}
\label{tp}
\mathcal{W}&=& - \frac{\Delta^2}{g} \\
\nonumber
&+&  \sum_{k} \Big[ \xi_{\mathbf{k}} -E_k-T \ln(e^{-\beta {\cal E}_{\mathbf{k}\uparrow} }+1)-T \ln(e^{-\beta {\cal E}_{\mathbf{k}\downarrow}}+1)\Big],
\end{eqnarray}
where we are considering $\Delta$ as real and $\beta=1/T$. We have also defined $\xi_{\mathbf{k}}= e_k - \mu$ as the single particle dispersion relation, with $E_k=\sqrt{ \xi_{\mathbf{k}}^2+\Delta^2 }$, and ${\cal E}_{\mathbf{k} \uparrow,\downarrow}= E_k \pm h$ are the quasiparticle excitations.

\section{Calculation of the Induced Interaction in the presence of impurities}
\label{Ind} 

The standard calculations of the GMB corrections are usually done presuming that the medium is clean. However, this is not always the case, particularly for superconductors for which impurities and defects are facilely present. An immediate consequence of impurities is that it may cause a finite lifetime for the quasiparticles. A concrete example happens in quasi-one-dimensional organic superconductors, in which lifetime effects emerge due to the existence of nonmagnetic impurities or defects~\cite{Ardavan}. The issue of impurities can make the effect of particle-hole fluctuations more involved, and therefore it is worth receiving the careful analysis we employ here.

The particle-hole fluctuations (or polarization function) $\chi_\text{ph}$ is found by considering a finite lifetime of the quasi-particle states in the momentum representation, i.e., by means of the usual procedure of adding a finite imaginary part to the bare Green's function. This means that it is assumed that the interaction of the fermions with impurity atoms is taken into account in $\mathcal{G}_0^\sigma(K)$~\cite{Abrikosov,Leung}, that is $\mathcal{G}_0^\sigma(K) \to \mathcal{G}_0^\sigma(K) = 1/(i\omega_l-\xi_{\vec{k}\sigma}+i\gamma\,\mathrm{sgn}(\xi_{\vec{k}\sigma}))$. For weak nonmagnetic impurities in the Born approximation that we consider here, one has $\gamma = 1/\tau = nu^2$, where $\tau$ is the mean free time due to elastic impurity scattering, $n=N/V$ is the impurity density, with $N$ being the number of impurities, $V$ the volume, and $u$ is the impurity scattering strength. See, e.g., Ref.~\cite {Abrikosov} for more information.

The authors of Ref.~\cite{Norman} used a model developed for copper oxides to investigate the behavior of the critical temperature $T_c$, normalized by the critical temperature in the absence of impurities $T_{c0}$ vs the bare impurity scattering rate $\tau^{-1}$ in meV for both weak- (inelastic scattering disregarded) and strong- (inelastic scattering regarded) coupling cases. For weak-coupling the range of $\gamma$ is from $0$ to $10$ and for strong-coupling $\gamma$ varies from $0$ to $30$. They found analytically and numerically the suppression of $T_c$ induced by structureless impurities treated within the Born approximation. This strong suppression of $T_c$ by nonmagnetic (as well as by magnetic) impurities has been confirmed experimentally further in, for example, Ref.~\cite{Mackenzie}.

Thus, the derivation of the GMB correction is performed in the presence of the spectral broadening. For the nonmagnetic impurities that we consider, alterations to the real part of the quasiparticle Green's function, if any, can be safely absorbed into the chemical potential. Together with the approximation of the imaginary part by a constant width parameter $\gamma$, weak nonmagnetic impurities satisfy the Anderson's theorem in the BCS limit \cite{Anderson,Note3,Chen-Schrieffer}.

This approach is quite similar to the one used to investigate the effects of nonmagnetic impurities in quasi-one dimensional imbalanced Fermi gases \cite{Continentino}, and also in two \cite{Mineev} and three \cite{Takada_1970} dimensional FFLO superconductors~\cite{FF,LO}.\\
\\ 
{\bf  The induced interactions}\\
\\

\begin{figure}
\begin{centering}
\includegraphics[clip,width=6cm]{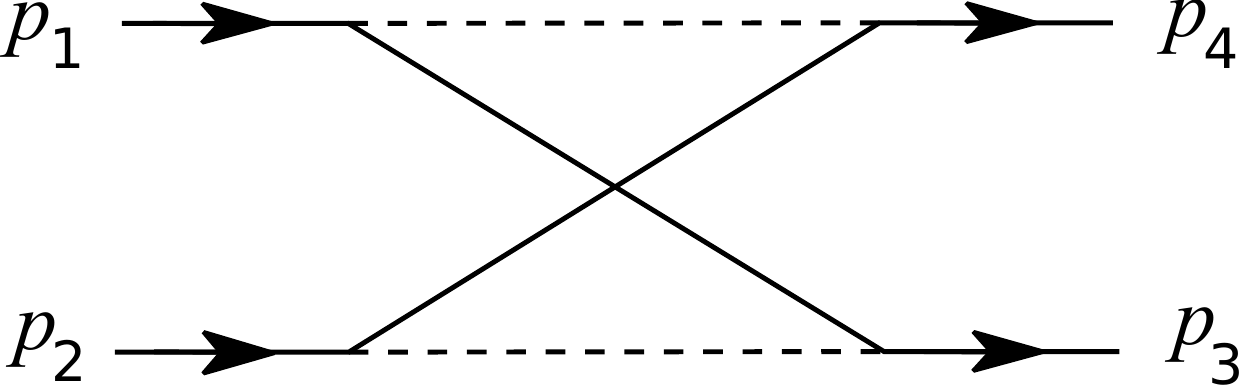}
\par\end{centering}
\caption[]{\label{LOD} The lowest-order diagram illustrating the induced interaction $U_{\text{ind}}( p_1, p_4)$. The solid and dashed lines describe fermion propagators and the bare interaction $g$ between fermions, respectively.}
\end{figure}

As GMB first realized, the interaction between two fermions in the medium is affected by the other fermions, leading to the screening of the effective interaction. The (lowest order) induced interaction was considered by GMB in the BCS limit by the second-order perturbation theory, in the calculation of $T_c$ of a dilute Fermi gas~\cite{Gorkov}. For a typical scattering process with $p_1+p_2\rightarrow p_3+p_4$, the induced interaction for the diagram depicted in Figure~\ref{LOD} is given by~\cite{Pethick00}

\begin{eqnarray}
\label{induced-weak}
U_{\text{ind}}( p_1, p_4)= -g^2 \chi_\text{ph}(p_1-p_4),
\end{eqnarray}
where $p_{i}=(\vec{k}_{i}, \omega_{l_i})$ is a four vector. 


The GMB correction can be extended to the strong interaction region~\cite{Yu10,Qijin}, in which case, to obtain an expression for the induced interaction, one considers the contribution of an infinite particle-hole ladder series that should replace the bare interaction $g$. The summation of this series provides the T-matrix in the particle-hole channel~\cite{Qijin},

\begin{eqnarray}
\label{tladder}
t_{ph}( p_1, p_4) &=& \frac{1}{g^{-1}+ \chi_\text{ph}( p_1, p_4)} = U_\text{eff}( p_1, p_4).
\end{eqnarray}
Taking into account the induced interaction, the effective pairing interaction $U_\text{eff}( p_1, p_4) \equiv U_\text{eff}$ between atoms with different spins is given by $U_\text{eff} = t_{ph}( p_1, p_4)= g + U_{\text{ind}}( p_1, p_4)$, then

\begin{eqnarray}
\label{induce}
U_{\text{ind}}( p_1, p_4) &=& U_\text{eff} - g\\
\nonumber
&=& \frac{g}{1+g \chi_\text{ph}( p_1, p_4)} - g \\
\nonumber
&=& -\frac{g^2 \chi_\text{ph}( p_1, p_4)}{1+g \chi_\text{ph}( p_1, p_4)}.
\end{eqnarray}
Notice that the leading term of a Taylor expansion of the denominator of the third line of Eq.~(\ref{induce}) in powers of $g \chi_\text{ph}$ is Eq.~(\ref{induced-weak}), originally considered by GMB~\cite{Gorkov}.

The polarization function $\chi_\text{ph}(p')$ is given by
\begin{eqnarray}
  \chi_\text{ph}(p')&=& \sum_p
                   \mathcal{G}_{0}^{\downarrow}(p)\mathcal{G}_{0}^{\uparrow}(p+p')\nonumber\\
               &=&\int {{\rm{d}}^3k \over (2\pi)^3} \frac{f_{{\vec k}}^{\downarrow}-f_{{\vec k}+{\vec q}}^{\uparrow}} {i\Omega_l+\xi_{{\vec k},\uparrow}-\xi_{{\vec k}+{\vec q},\downarrow}},
\label{chi0}
\end{eqnarray}
where $f_{{\vec k}}^\sigma\equiv f(\xi_{\vec{k},\sigma})=1/[exp(\beta \xi_{\vec{k},\sigma})+1]$ is the Fermi distribution function, and $p'=(\vec{q}, \Omega_{l})$ is a four-vector with bosonic Matsubara frequency $\Omega=2\pi l T$, with $l$ an integer. A simple inspection shows that the effective interaction $U_\text{eff}$ depends on the momentum and frequency. After performing the analytical continuation $\omega_0 \to \omega_0 + i\gamma$~\cite{Abrikosov,Leung}, the static polarization function is calculated in Appendix A, and we find
\begin{eqnarray}
\label{generalizedL}
\tilde{\chi}_\text{ph}(x,y,\tilde \gamma) = - N(0) L(x,y,\tilde \gamma),
\end{eqnarray}
where $x\equiv \frac{q}{2 k_\text{F}}$, $y\equiv \frac{h}{E_F}$, and $\tilde \gamma \equiv \frac{\gamma}{E_F}$, and $N(0)=mk_F/(2\pi^2)$ is the usual density of states at the Fermi surface for a single spin species.

\section{Effects of induced interactions corrections to the transition temperature of a balanced Fermi gas}
\label{Tc}

In this section we show that, indeed, the transition temperature can be strongly affected by impurity effects \cite{Agterberg}. As we mentioned earlier, we shall only consider nonmagnetic weak impurities in the Born approximation~\cite{Che_Born} which, as explained above, mainly lead to a finite spectral broadening $\gamma$ for the fermions.

Notice also that in the $\tilde \gamma \to 0$ and $y \to 0$ limits in Eqs.~\ref{chiph-full-1} and~\ref{chiph-full-2}, $k_\text{F}^{\downarrow}=k_\text{F}^{\uparrow}=k_\text{F}$, such that $\tilde{\chi}_\text{ph}^{\downarrow}(x,0,0) = \tilde{\chi}_\text{ph}^{\uparrow}(x,0,0) \equiv \tilde{\chi}_\text{ph}(x)/2$ and we obtain the well-known (balanced and without impurities) textbook result,
\begin{eqnarray}
\nonumber
\tilde{\chi}_\text{ph}(x) &=&- \frac{m}{ 4 \pi^2 q} \left[  k_\text{F}  q -   \left( k_\text{F}^2- \frac{q^2}{4}  \right ) \ln \left| \frac{q^2 - 2q k_\text{F} }{q^2 + 2q k_\text{F}} \right|   \right]\\
&=& -N(0)  L(x),
\label{chiph-9}
\end{eqnarray}
where $ L(x)\equiv L(x,0)$ is the standard (static) Lindhard screening function,
\begin{eqnarray}
\label{chiph9-2}
 L(x)= \frac{1}{2} -\frac{1}{4x}(1-x^2) \ln \left|\frac{1-x}{1+x} \right|.
\end{eqnarray}

Conservation of the (total) momentum requires that in the scattering event $\vec{k}_1+\vec{k}_2=\vec{k}_3+\vec{k}_4$, with $\vec{k}_1=- \vec{k}_2$ and $\vec{k}_3=-\vec{k}_4$. $q$ is the modulus of $\vec{k}_1+\vec{k}_3$ i.e., $q=\sqrt{(\vec{k}_1+\vec{k}_3).(\vec{k}_1+\vec{k}_3)} = \sqrt{\vec{k}_1^2+\vec{k}_3^2+  2\vec{k}_1. \vec{k}_3}=\sqrt{\vec{k}_1^2+\vec{k}_3^2+ 2|\vec{k}_1||
  \vec{k}_3|\cos \phi}$, where $\phi$ is the angle formed by $\vec{k}_1$ and $\vec{k}_3$. Since the scattering particles are at the Fermi surface, we have $|\vec{k}_1|=|\vec{k}_3|=k_\text{F}=\sqrt{2m\mu}$. Then, $q=k_\text{F}\sqrt{2(1+\cos \phi)}$, which implies in $x=\sqrt{2(1+\cos \phi)}/2$, and this imposes the $x$ domain as $0 \leq x \leq1$. Taking the average of ${\chi}_\text{ph}(x)$ we are left with the well known GMB result $\langle\tilde{\chi}_\text{ph}(x)\rangle = - N(0) (1+2\ln 2)/3 = - N(0) \ln(4e)^{1/3}$.

To obtain the GMB correction to $T_c$ of a balanced Fermi gas in the presence of impurities we simply have to impose $y=0$ in Eqs.~\ref{chiph-full-1} and~\ref{chiph-full-2}, which yields $\tilde{\chi}_\text{ph}^{\downarrow}(x,\tilde \gamma) = \tilde{\chi}_\text{ph}^{\uparrow}(x,\tilde \gamma) \equiv \tilde{\chi}_\text{ph}(x,\tilde \gamma)$ then,

\begin{widetext}
\begin{eqnarray}
\label{chiph-full-3}
\tilde{\chi}_\text{ph}(x,\tilde \gamma) = - N(0) 2 L(x,\tilde \gamma) = &-& N(0) \biggl[ \frac{1}{2} -  \frac{1}{4x} \left[ 1 - x^2 + \left( \frac{\tilde \gamma }{4 x}\right)^2 \right] \frac{1}{2} \ln \left( \frac{(1 - x)^2 + \left( \frac{\tilde \gamma }{4 x}\right)^2}{(1 + x)^2 + \left( \frac{\tilde \gamma }{4 x}\right)^2} \right)  \\
\nonumber
&-& \frac{1}{2} \left( \frac{\tilde \gamma }{4 x}\right)  \left[ \arctan \left( \frac{4 x -4x^2 }{ \tilde \gamma} \right) + \arctan \left( \frac{4 x +4x^2 }{ \tilde \gamma}  \right) \right] \biggr].
\end{eqnarray}
\end{widetext}
Notice that the limit $\tilde \gamma \to 0$ in $L(x,\tilde \gamma)$ above, immediately gives the standard $L(x)$ from Eq.~(\ref{chiph9-2}).

\begin{figure}
\begin{centering}
\includegraphics[clip,width=8cm]{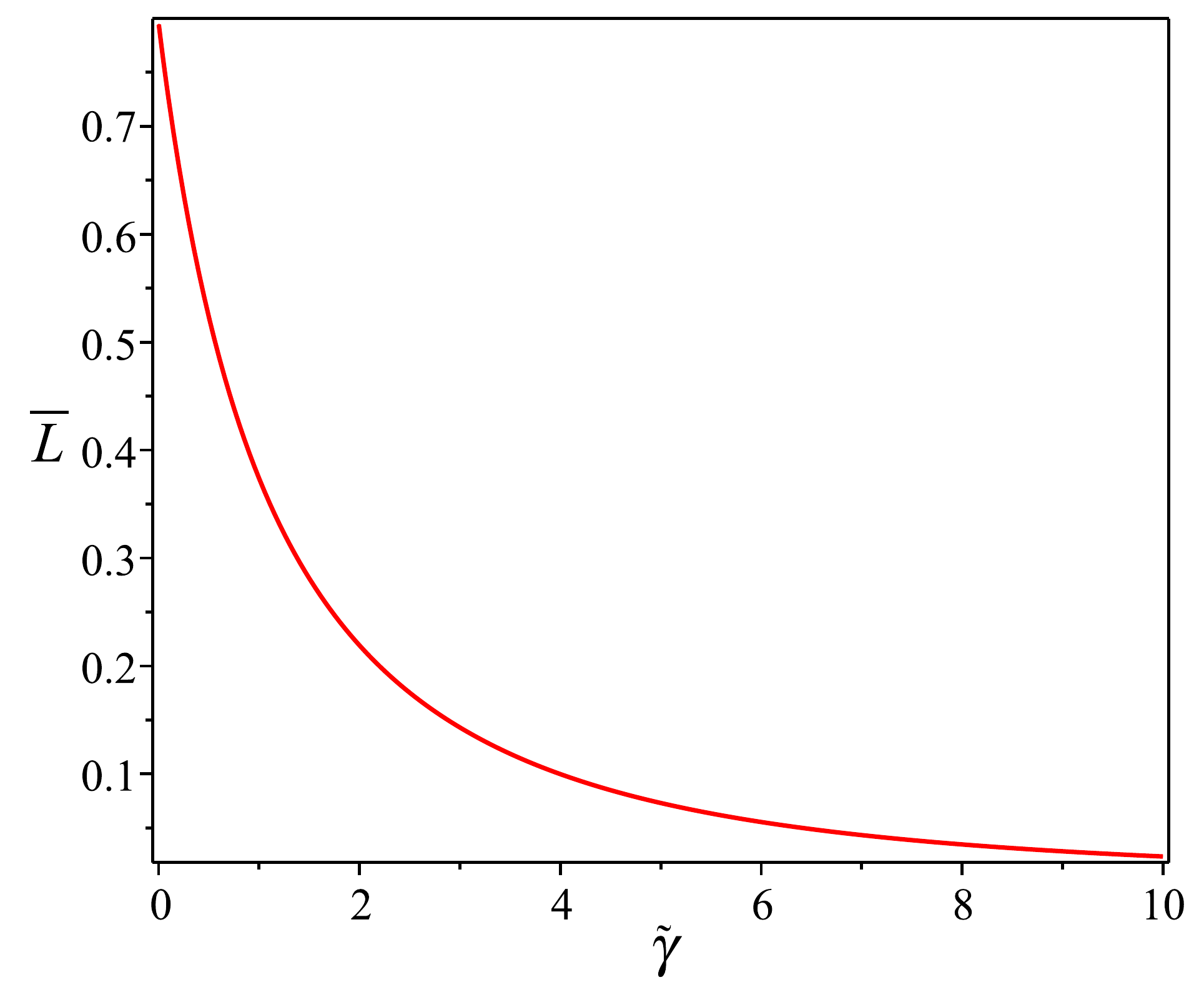}
\par\end{centering}
\caption[]{\label{Balanced} The averaged generalized Lindhard function $\bar L(\tilde \gamma)$, for balanced Fermi systems (i.e., $y=h=0$), versus the impurity parameter $\tilde \gamma$. The value $\bar L(\tilde \gamma=0) = 0.7954$ corresponds to the standard (clean) value.}
\end{figure}

In Figure~\ref{Balanced} the function $\bar L(\tilde \gamma)$ (the averaged $L(x,\tilde \gamma)$ from Eq.~(\ref{chiph-full-3})) is shown as a function of the non-dimensional impurity parameter $\tilde \gamma$. The result shows that $\bar L(\tilde \gamma)$ vanishes asymptotically i.e., $\bar L(\tilde \gamma)$ decreases and approaches $0$ as $\tilde \gamma$ gets larger. The value $\bar L(\tilde \gamma=0) = 0.7954$ corresponds to the standard averaged Lindhard function $\bar L(x)$ in the clean limit~\cite{Pethick00,Qijin}.

To obtain a thermal gap equation we extremize the grand thermodynamic potential in Eq.~(\ref{tp}) with respect to $\Delta$. After going from summations to integrals we find the critical temperature $T_c$ as the one such that $\Delta(T_c) = 0$. Then we have

\begin{eqnarray}
\label{ap9}
\frac{1}{g} &+& \langle \tilde{\chi}_\text{ph}(x,y,\tilde \gamma) \rangle \\
\nonumber
&+&\int \frac{d^3 k}{(2 \pi)^3}  \frac{1}{2 \xi_{\mathbf{k}}} \left[ 1-f({\xi_{\mathbf{k}\uparrow}}) - f({\xi_{\mathbf{k}\downarrow}}) \right] =0,
\end{eqnarray}
where $\xi_{\mathbf{k}}=k^2/2m-\mu$. In Eq.~(\ref{ap9}) the GMB correction has been taken into account through the effective coupling constant. The equation above is equivalent to the divergence of the T-matrix, which is given by
\begin{equation}
  t^{-1}(\mathbf{q},\Omega=0)=\left(\frac{1}{g} + \langle \tilde{\chi}_\text{ph}(x,y,\tilde \gamma) \rangle \right)+\chi({\vec{q}},\Omega=0)=0,
  \label{eq:22}
\end{equation}
where $\chi({\vec{q}},\Omega)$ is the bare pair susceptibility, with $({\vec{q}},\Omega)$ being the four momentum of pairs. The above equation can be obtained by replacing $g$ by $U_\text{eff}$, as given in Eq.~\ref{tladder}. Equation~(\ref{eq:22}) was shown to be correct also when the T-matrix includes the particle-hole channel in a self-consistent approximation \cite{Qijin}. Setting $y=0$ in Eq.~(\ref{eq:22}) yields a GMB corrected solution

\begin{equation}
\label{hcrit}
T_c(\tilde \gamma) = T_{c,0}^\text{MF} e^{-\bar L(\tilde \gamma)},
\end{equation}
where $T_{c,0}^\text{MF}=(8/\pi) e^{\gamma-2} \mu e^{-\pi/2k_F|a|}$ ($\gamma = 0.5772$ is Euler's constant) is the MF result without the GMB corrections and in the clean limit. 

It is well known that the BCS transition temperature of a balanced Fermi gas is modified due to the particle-hole channel effect (or GMB correction) as~\cite{Pethick00,Qijin}
\begin{eqnarray}
T_c^\text{GMB} &=& \frac{T_{c,0}^\text{MF}}{(4e)^{1/3}},
\label{GMBgap}
\end{eqnarray}
which corresponds precisely to $T_c(y=\tilde \gamma=0)$ from Eq.~(\ref{hcrit}).

In the manner the calculations have been done until here, as became clear during the steps, the BCS instability occurring in the particle-particle channel and the polarization effects of the medium occurring in the particle-hole channel got totally disentangled from each other, thus recovering the original GMB result~\cite{Pisani}. The treatment developed in Ref.~\cite{Pisani} for the GMB correction of a {\it balanced and clean} Fermi gas has clear advantages over previous ones on the same subject~\cite{Yu09,Ruan}. This is due to the fact that they considered the GMB correction not only in the BCS limit but also across the BCS-BEC crossover, albeit employing a sequence of (justified) necessary approximations to achieve analytical results.

The employment of the T-matrix approximation deserves some comments at this point. It is widely accepted that the ladder sum in the (strong-coupling) particle-particle channel for a Fermi gas at unitarity is necessary, but is not sufficient, since it is not the complete set of all possible diagrams in perturbation theory~\cite{Hu}. In order to compare theoretical predictions of thermodynamic properties of strongly interacting Fermi gases, such as the energy $E(T)$ and entropy $S(T)$, with experimental measurements, several versions of the T-matrix approximation have been used. The distinct versions give conflicting predictions, depending on which Green's functions $G_0$ (non-interacting Green's function) or $G$ (fully dressed interacting Green's function) are chosen to enter the pair-propagator~\cite{Hu,gen}. Notice that even other methods used to build the phase diagram, as scaling functions within a Luttinger-Ward approach, rely on the Green's function that solves the self-consistent Dyson equation~\cite{Lang}.

In order to improve the non-selfconsistent T-matrix approximation~\cite{gen} to obtain the critical temperature $T_c$ throughout the BCS-BEC crossover, Pisani and co-workers~\cite{Pisani} included the Popov and GMB corrections into the Thouless criterion, which amounts to rewrite Eq.~(\ref{eq:22}) as

\begin{equation}
 \frac{m}{4\pi a} + R_{pp}(q=0) + \Sigma^B=0,
  \label{Pis1}
\end{equation}
where $R_{pp}(q=0)$ is the regularized particle-particle bubble (equivalent to a regularized $\chi({\vec{q}},\Omega)$, defined below Eq.~(\ref{eq:22}), as we will use in the next Sections), and $\Sigma^B$ represents the bosoniclike self-energies. In order to obtain the critical temperature all through the BCS-BEC crossover as a function of $1/k_Fa$, various approximations have been implemented by taking distinct $\Sigma^B$, namely $\Sigma^B_{GMB}$, $\Sigma^B_{Popov}$ or $\Sigma^B_{GMB} + \Sigma^B_{Popov}$. Interestingly, the several approximations give a maximum $T_c/T_F$, but each of them at a different value of the ``interaction'' belonging to the interval $-1 \lesssim 1/k_Fa \lesssim +1$. At weak-coupling $\Sigma^B_{GMB}$ reduces to our $\chi_{ph}$ in the limit $y=0$ and $\tilde \gamma=0$. In the BEC side, at the strong limit of the GMB, they found a final expression for the bosoniclike self-energy that, to the leading order in the small parameter $k_Fa$, yields $\Sigma^B_{GMB} \simeq \frac{16 {\it \tilde I}}{\pi^2} \frac{m k_F}{\pi^2}(k_Fa)^2$, where $ {\it \tilde I} = 0.25974$~\cite{Pisani}. 

In the BEC (strong-coupling) limit ($1/k_Fa \gg 1$) all alternative T-matrix approaches reproduce the known value of the critical temperature for a condensate of noninteracting composite bosons (i.e., made up of fermion pairs), with mass $m_B=2m$~\cite{Pini}:

\begin{equation}
 T_c^{BEC}= \frac{2\pi}{\zeta(3/2)^{2/3}} \frac{{n_B}^{2/3}}{m_B}.
  \label{TBEC}
\end{equation}
Far in the BEC regime all fermions form pairs, therefore $n_B=n/2$. At fixed fermionic density $E_F = \frac{(6\pi^2 n)^{2/3}}{2m}$ , then we have

\begin{equation}
 T_c^{BEC}=\left( \frac{\sqrt{2}}{3 \sqrt{\pi}\zeta(3/2)}\right)^{2/3}  \simeq 0.218 E_F,
  \label{TBEC2}
\end{equation}
which is the condensation temperature of an ideal Bose gas in three dimensions. In the above equation we used that $\zeta(3/2)=2.612$, with $\zeta(z)$ the Riemann Zeta function.

\section{Effects of induced interactions corrections to the tricritical point of an imbalanced Fermi gas}
\label{TcP}

We firstly investigate the case of the clean imbalanced Fermi gas. To this aim, we simply have to set $\tilde \gamma=0$ in Eq.~\ref{generalizedLFinal}.

\begin{figure}
\begin{centering}
\includegraphics[clip,width=8cm]{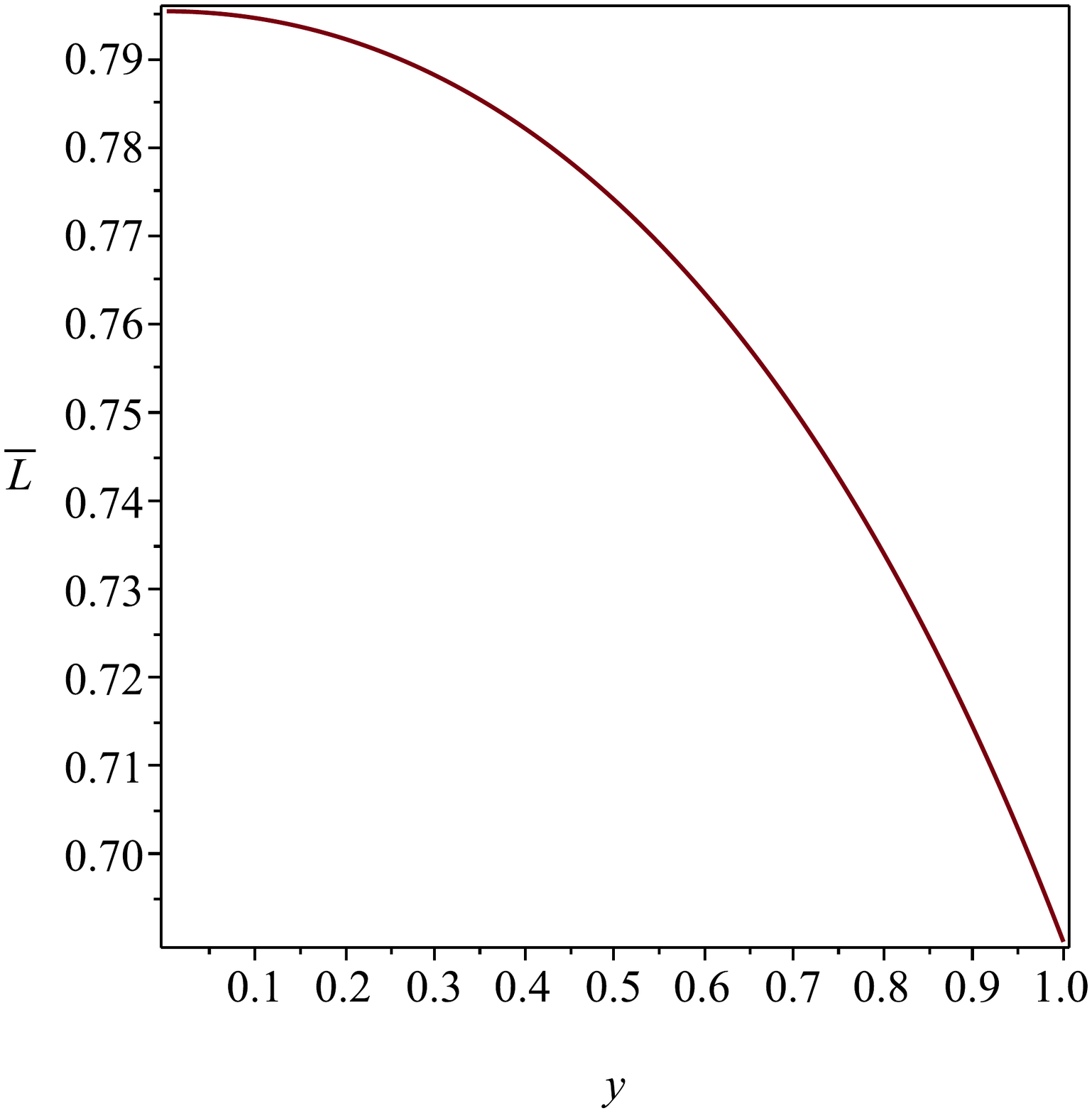}
\par\end{centering}
\caption[]{\label{L} Behavior of $\bar L(y)$ as a function of the imbalance $y$ for a clean imbalanced 3D homogeneous Fermi gas.}
\end{figure}

The effective induced interaction is obtained by taking the average of the polarization function $\chi_\text{ph}(x,y)$ over the Fermi surface, which means an average of the angle
$\phi$~\cite{Pethick00,Qijin,Petrov,Baranov2,Yu09,Torma,Yu10},
\begin{eqnarray}
\langle\tilde{\chi}_\text{ph}(x,y)\rangle &=& \frac{1}{2} \int_{-1}^{1} \mathrm{d} \cos\phi ~\tilde{\chi}_\text{ph}(x,y)\nonumber\\
&=& - N(0) \left[  \frac{1}{2} \int_{-1}^{1} \mathrm{d} \cos\phi ~ L(x,y)\right] \nonumber\\
&\equiv& - N(0) \bar L(y),
\label{FirstAverage}
\end{eqnarray}
where Eq.~\ref{generalizedLFinal} has been used. The quantity $\bar L$ characterizes the magnitude of GMB corrections in the presence of population imbalance. We show in Figure~\ref{L} the behavior of $\bar L(y)$ as a function of the (non-dimensional) imbalance $y$. In the limit $y\rightarrow 0$, we have precisely $\bar L(0) =(1+2\ln 2)/3 = 0.7954 $ found above, as given in Ref.~\cite{Qijin} and several other papers as, for instance, \cite{Yu10} for the balanced case. As $y$ increases from 0 to 1, $\bar L(y)$ decreases from $0.7954$ to $0.69$, showing that the particle-hole fluctuation is weakened as a consequence of the Fermi surface mismatch caused by an increase of the population imbalance. This result is the same as the one found in Ref.~\cite{Yu10}.

To obtain the GMB correction to the mean-field tricritical point in the presence of impurities, differently from what was done above, we use Eq.~\ref{generalizedLFinal} with a finite $\tilde \gamma$ and follow Ref.~\cite{Resende} to define

\begin{eqnarray}
\label{ap10}
{\mathcal F}(T,h)&=&- \frac{1}{g} - \langle \tilde{\chi}_\text{ph}(x,y,\tilde \gamma) \rangle\\
\nonumber
&-& \int \frac{d^3 k}{(2 \pi)^3}  \frac{1}{4 \xi_{\mathbf{k}}} \left[ \tanh \left({\frac{\beta \xi_{\mathbf{k}\uparrow}}{2}} \right) + \tanh \left({\frac{\beta \xi_{\mathbf{k}\downarrow}}{2}} \right) \right],
\end{eqnarray}
where $\langle ~\rangle$ is the angle average of $ \tilde{\chi}_\text{ph}(x,y,\tilde \gamma) $ given by Eq.~(\ref{generalizedLFinal}) (please, see below Eq.~(\ref{ap16})).

In order to make contact with experiments, we use the Lippmann-Schwinger equation to trade the bare coupling constant $g$ by the s-wave scattering length $a$

\begin{eqnarray}
\label{ap11}
\frac{1}{g} = \frac{m}{4 \pi a} - \int \frac{d^3k}{(2\pi)^3} \frac{1}{2e_k},
\end{eqnarray}
where $e_k=k^2/2m$. Then we have,

\begin{eqnarray}
\label{ap12}
&&{\mathcal F}(T,h)= - \frac{m}{4 \pi a} - \langle \tilde{\chi}_\text{ph}(x,y,\tilde \gamma) \rangle\\
\nonumber
&-& \int \frac{d^3 k}{(2 \pi)^3} \left[ \frac{1}{4 \xi_{\mathbf{k}}} \left( \tanh \left({\frac{\beta \xi_{\mathbf{k}\uparrow}}{2}} \right) + \tanh \left({\frac{\beta \xi_{\mathbf{k}\downarrow}}{2}} \right) \right) -\frac{1}{2e_k}\right].
\end{eqnarray}

We now define ${\mathcal G} \equiv 1/\Delta^2 \partial^2 \Omega/\partial \Delta^2|_{\Delta=0}$, which yields

\begin{eqnarray}
\label{ap13}
{\mathcal G}(T,h) &=&  \int \frac{d^3 k}{(2 \pi)^3}  \frac{\beta}{2 \xi_{\mathbf{k}}^2} \big[\frac{1-f_{{\vec k}}^{\uparrow}-f_{{\vec k}}^{\downarrow}}{\beta \xi_{\mathbf{k}}} \\
\nonumber
&+&  \sum_{\sigma} f_{{\vec k}}^{\sigma}(f_{{\vec k}}^{\sigma}-1)  \big].
\end{eqnarray}

The functions ${\mathcal F}(T,h)$ and ${\mathcal G}(T,h)$ correspond to $\alpha_{GL}$ and $\beta_{GL}$, the first and second coefficients of the expansion of the free energy in terms of the gap parameter, according to the Landau theory of phase transitions. Thus, by setting $\alpha_{GL}=\beta_{GL}=0$, one finds $h_{tc}$ and $T_{tc}$ of the tricritical point (TCP). Below this point, i.e., at low temperatures, the transition is of first-order and the critical temperature has to be found by properly equating the thermodynamic potentials $\mathcal{W}$ of the superfluid and normal phases, namely $\mathcal{W}^S(T_c,\Delta=\Delta_0)=\mathcal{W}^N(T_c,\Delta=0)$, where $\Delta_0$ is the minimum of $\mathcal{W}^S$~\cite{Chapter,JSTAT1}. From Eq.~(\ref{tp}) we find,

\begin{eqnarray}
\label{equaltp}
&-&  \left[ \frac{m}{4 \pi a} + \langle \tilde{\chi}_\text{ph}(x,y,\tilde \gamma) \rangle - \sum_{k} \frac{1}{2e_k}  \right] \Delta_0^2 \\
\nonumber
&+&  \sum_{k} \Big[ \xi_{\mathbf{k}} -E_k-T \ln(e^{-\beta {\cal E}_{\mathbf{k}\uparrow} }+1)-T \ln(e^{-\beta {\cal E}_{\mathbf{k}\downarrow}}+1)\Big]\\
\nonumber
&=& \sum_{k} \Big[ \xi_{\mathbf{k}} -|\xi_{\mathbf{k}}|-T \ln(e^{-\beta {\cal E}_{\mathbf{k}\uparrow}^0 }+1)-T \ln(e^{-\beta {\cal E}_{\mathbf{k}\downarrow}^0}+1)\Big],
\end{eqnarray}
where ${\cal E}_{\mathbf{k} \uparrow,\downarrow}^0 \equiv {\cal E}_{\mathbf{k} \uparrow,\downarrow}(\Delta=0)= |\xi_{\mathbf{k}}| \pm h$. From the above equation, for each imbalance $y$ belonging to the interval $y_{c} < y < y_{CC}$ one finds the corresponding $\tilde T_c$ of the first-order curve in the $(y,\tilde T_c)$-phase diagram. Here $y_{c}$ is the imbalance of the tricritical point $(y_{c},\tilde T_c)$. The MF Chandrasekhar-Clogston (CC) first-order transition limit $y_{CC}$ is obtained by taking the zero temperature limit of Eq.~(\ref{equaltp}), such that the equality of the energies of the {\it balanced} BCS state and imbalanced normal state gives

\begin{eqnarray}
\label{CC}
&&  \sum_{k} \Big[ \xi_{\mathbf{k}} -E_k \Big] -\frac{\Delta_0^2}{g}\\
\nonumber
&=& \sum_{k} \Big[ (\xi_{\mathbf{k}} + h) \Theta(-\xi_{\mathbf{k}} - h) + (\xi_{\mathbf{k}} - h) \Theta(h -\xi_{\mathbf{k}} ) \Big],
\end{eqnarray}
which yields $E(0) - N(0) \Delta_0^2/2 = E(0) - N(0) h^2$ in the limit $h\ll \mu$, where $E(0) \equiv E(\Delta=0)$ is the energy of the balanced normal state, and from this one obtains $h_{CC}=\Delta_0/\sqrt{2}$~\cite{Chandrasekhar,Clogston}, which is exponentially small. (Please, see the text on Figure~\ref{PD} below).

The two functions that have to be solved self-consistently are

\begin{widetext}
\begin{eqnarray}
\label{ap14}
\frac{m}{4 \pi a} + \langle \tilde{\chi}_\text{ph}(x,y,\tilde \gamma) \rangle + \int \frac{d^3 k}{(2 \pi)^3} \left[ \frac{1}{4 \xi_{\mathbf{k}}} \left( \tanh \left({\frac{\beta \xi_{\mathbf{k}\uparrow}}{2}} \right) + \tanh \left({\frac{\beta \xi_{\mathbf{k}\downarrow}}{2}} \right) \right) -\frac{1}{2e_k}\right]=0,
\end{eqnarray}

\begin{eqnarray}
\label{ap15}
\int \frac{d^3 k}{(2 \pi)^3}  \frac{\beta}{2 \xi_{\mathbf{k}}^2} \left[\frac{1-f_{{\vec k}}^{\uparrow}-f_{{\vec k}}^{\downarrow}}{\beta \xi_{\mathbf{k}}} +  \sum_{\sigma} f_{{\vec k}}^{\sigma}(f_{{\vec k}}^{\sigma}-1)  \right]=0,
\end{eqnarray}
\end{widetext}
which in dimensionless variables read

\begin{widetext}
\begin{eqnarray}
\label{ap16}
\frac{1}{4 \pi k_F a} -\frac{ \bar L(y,\tilde \gamma)}{2\pi^2} + \int_0^Z \frac{z^2 dz}{2 \pi^2} \left\{ \frac{1}{2(z^2-1)} \left[ \tanh \left({\frac{\tilde\beta \xi_{\uparrow}}{2}} \right) + \tanh \left({\frac{\tilde\beta \xi_{\downarrow}}{2}} \right) \right] -\frac{1}{z^2} \right\} =0,
\end{eqnarray}
where $\bar L(y,\tilde \gamma) =  \frac{1}{2} \int_{-1}^{1} \mathrm{d} \cos\theta ~ L(x,y,\tilde \gamma)$, with $L(x,y,\tilde \gamma)$ given by Eq.~(\ref{generalizedLFinal}), in which $x\equiv \frac{q}{2 k_\text{F}}=\sqrt{2(1+\cos \phi)}/2$, $y\equiv \frac{h}{E_F}$, $\tilde \gamma \equiv \frac{\gamma}{E_F}$, $Z \equiv \Lambda/k_F$ is a cutoff that may be extended to infinity, so that no one of the results shall depend on it, $\tilde \beta \equiv 1/\tilde T = 1/(T/E_F)$, and $\xi_{\uparrow,\downarrow} \equiv \sqrt{(z^2-1)^2}\pm y = |z^2-1| \pm y$ or, in terms of the Fermi functions,

\begin{eqnarray}
\label{ap17}
\frac{1}{ k_F a} -\frac{2 \bar L(y,\tilde \gamma)}{\pi} + \frac{2}{\pi} \int_0^Z z^2 dz \left\{ \frac{1}{2(z^2-1)} \left[ 1-f^{\uparrow} - f^{\downarrow} \right] -\frac{1}{z^2} \right\} =0,
\end{eqnarray}
and

\begin{eqnarray}
\int_0^Z z^2 dz  \frac{1}{(z^2-1)^2}   \left\{ \frac{1}{2\tilde \beta (z^2-1)}\left[ \tanh \left({\frac{\tilde\beta \xi_{\uparrow}}{2}} \right) + \tanh \left({\frac{\tilde\beta \xi_{\downarrow}}{2}} \right) \right] - \frac{1}{4}  \sum_{\sigma} \rm{sech}^2 \left({\frac{\tilde\beta \xi_{\sigma}}{2}} \right)\right\}=0,~~\rm{or}\\
\nonumber
\end{eqnarray}

\begin{eqnarray}
\int_0^Z z^2 dz  \frac{1}{(z^2-1)^2}   \left\{ \frac{1-f^{\uparrow} - f^{\downarrow}}{\tilde \beta (z^2-1)} +  \sum_{\sigma} f^{\sigma}(f^{\sigma}-1) \right\}=0,
\label{ap18}
\end{eqnarray}
\end{widetext}
where $f^{\sigma} \equiv f(\tilde \beta \xi_{\sigma})=1/[exp(\tilde \beta \xi_{\sigma})+1]$. 

\begin{figure}
\centerline{\includegraphics[clip,width=4.0in]{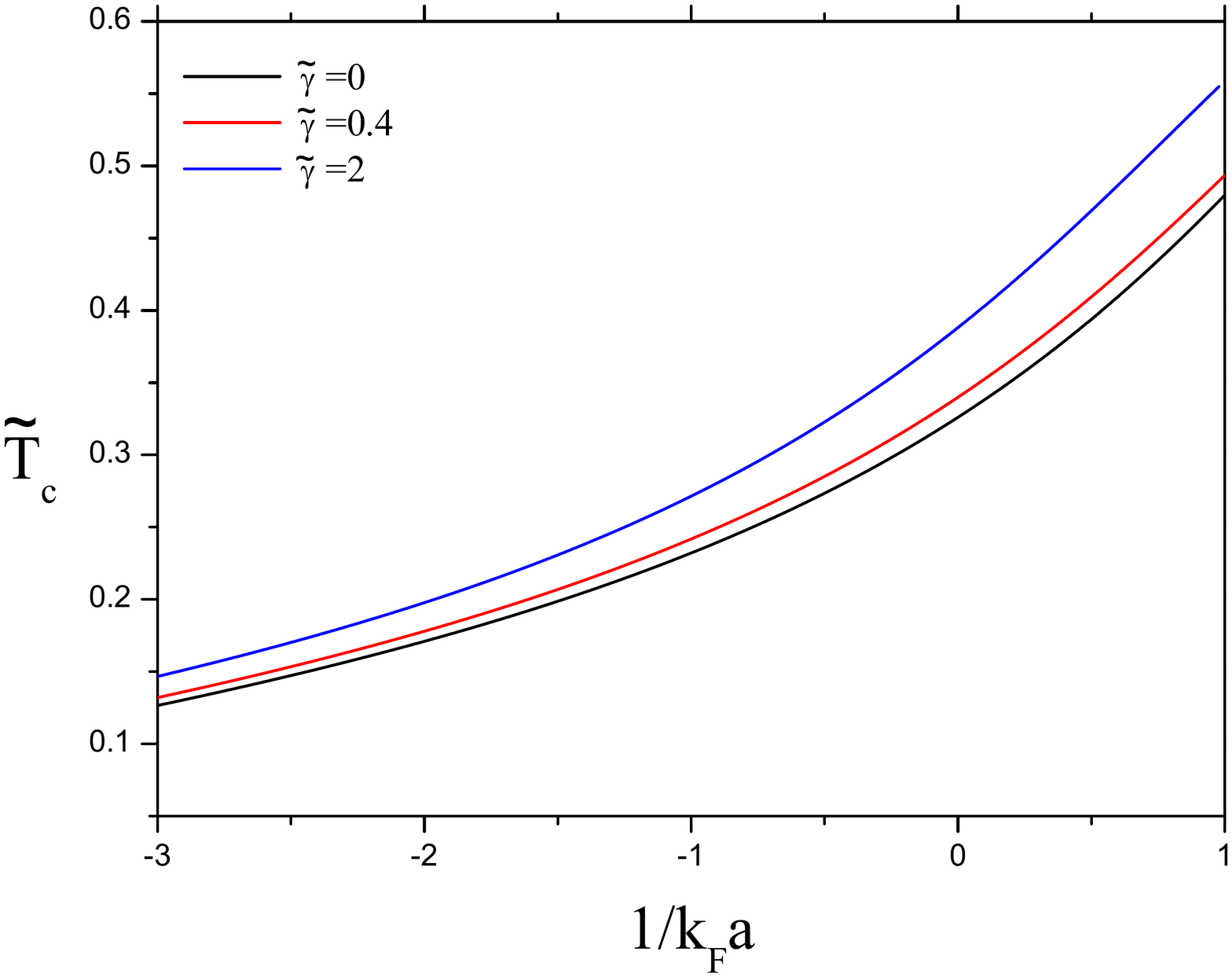}}
\caption{Non-dimensional critical temperature $\tilde T_c$ of the tricritical point as a function of the interaction parameter $1/k_Fa$ for various values of the impurity parameter $\tilde \gamma$.}
\label{Tcdelta}
\end{figure}

\begin{figure}
\centerline{\includegraphics[clip,width=4.0in]{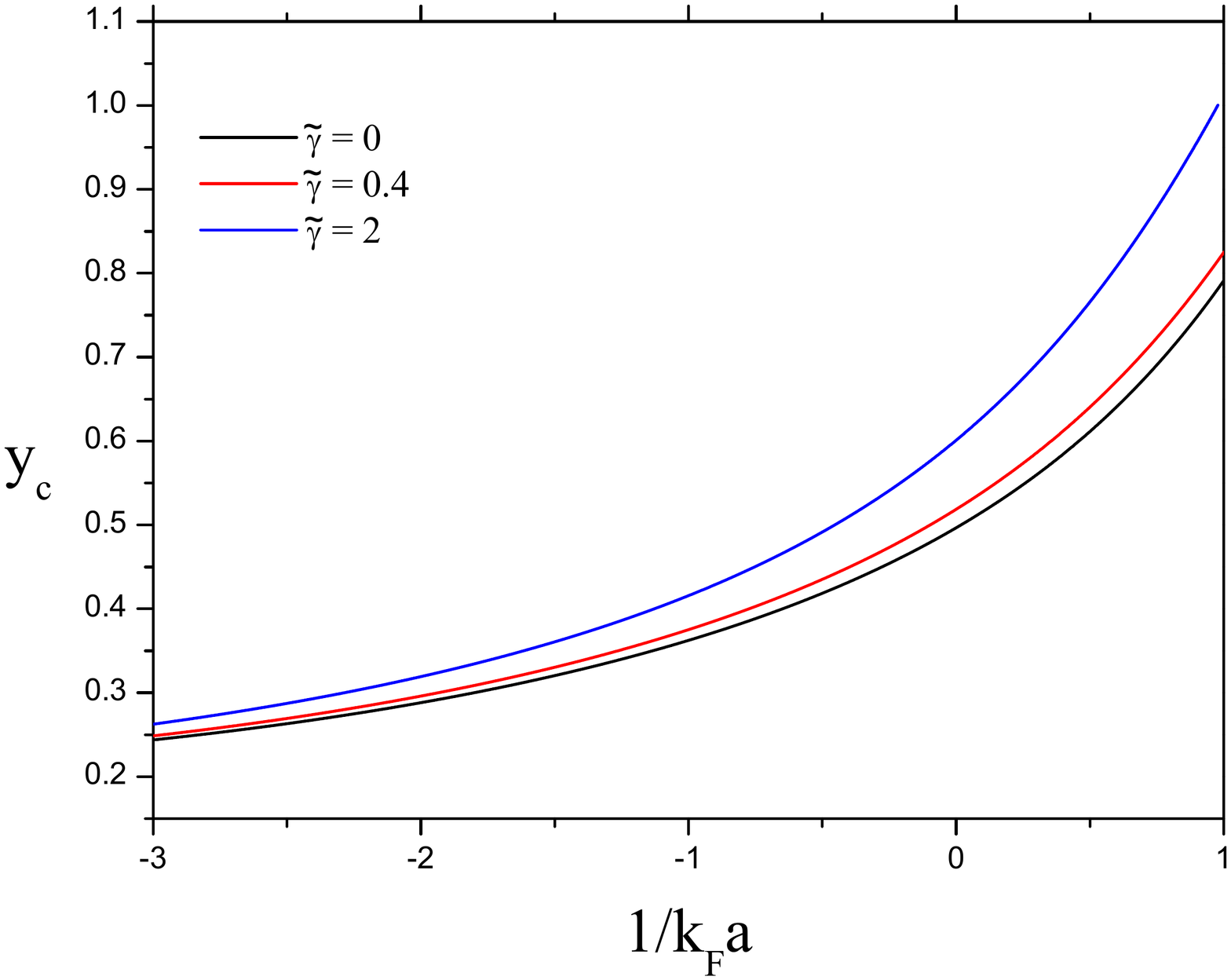}}
\caption{Behavior of the non-dimensional critical chemical potential difference $y_c$ of the tricritical point as a function of the interaction parameter $1/k_Fa$ for several values of the impurity parameter $\tilde \gamma$.}
\label{Ycdelta}
\end{figure}

In Figures~\ref{Tcdelta} and~\ref{Ycdelta} we show the numerical solutions of Eqs.~(\ref{ap17}) and~(\ref{ap18}) as a function of $1/k_Fa$ for some values of the impurity parameter $\tilde \gamma$. Notice that the GMB correction is present only in the equation for $\alpha_{GL}=0$ (Eq.~(\ref{ap17})). However, since Eqs.~(\ref{ap17}) and~(\ref{ap18}) were solved self-consistently, the polarization corrections affect both $\tilde T_c$ and $y_c$ of the tricritical point. The bottom (black) curve for $\tilde T_c$ (and the same for $y_c$) as a function of $1/k_Fa$ is the MF corrected by the GMB corrections, in the clean limit ($\tilde \gamma=0$). Observe that, as expected, $\tilde T_c$ and $y_c$ increase with increasing $k_Fa$. For a fixed value of $1/k_Fa$, as $\tilde \gamma$ increases (from bottom to top), $\tilde T_c$ and $y_c$ also increase, approaching the MF values. Physically this happens because the scattering of the other fermions (responsible for the screening) with impurities in the medium, affects the screening of the interaction between the $\uparrow$ and $\downarrow$ fermions. Besides, as the results show, a high level of impurities can even nullify the GMB correction. 

Although we did not find in the literature $\tilde T_c$ and $y_c$ corrected by the GMB in the presence of impurities to compare with, we find that the results shown in Figures~\ref{Tcdelta} and~\ref{Ycdelta} agree qualitatively with the ones shown in Figure 2 of the work by Yu and Yin~\cite{Yu10}. In Fig.~2, (a) and (b), of Ref.~\cite{Yu10} they find that the tricritical polarization $p_t$ and reduced temperature $T_t$, respectively, of (clean) imbalanced Fermi gases are decreased (in comparison to the MF results) due to the GMB corrections, as we find here.

In Fig.~\ref{PD} we show the phase diagram in the $(y,\tilde T)$-plane of the homogeneous spin-imbalanced unitary Fermi gas, characterized by the divergence of the s-wave scattering length, which implies in $1/k_Fa=0$. The top curve is the standard MF (without the GMB correction) phase boundary between the SF and N phase. The point $(0,0.64)$ at the $\tilde T_c$ axes is the critical temperature of the (balanced) BCS case. The tricritical point at the end of the MF second-order curve is given by $(y_c,\tilde T_c) \approx (0.63,0.37)$, in agreement with previous findings (see, for instance, Refs.~\cite{Casalbuoni,Chevy2010,Gubbels,Boettcher,Strinati} for reviews). The tricritical point of the MF corrected by GMB second-order (bottom, black) curve is at $(y_c,\tilde T_c) \approx (0.50,0.32)$, which is also in agreement with experiments and the results obtained by means of different theoretical methods by several authors, see a detailed comparison in~\cite{Sylvain}. Above the bottom (black) line, we show the curves with $\tilde \gamma=0.5, 2.0~\rm{and}~6.0$, respectively. A first-order line (also in orange), obtained from the non-dimensional version (Eq.~(\ref{firstorder-1}) of appendix~(\ref{ApB})) of Eq.~(\ref{equaltp}), starts at the tricritical point of the MF curve and ends at the Chandrasekhar-Clogston (or Pauli) limit of critical polarization $(y_{CC},0)=(0.81,0)$~\cite{Chandrasekhar,Clogston} (pink dot at the $y_c$ axes). At this point, $h/\Delta_0=1/\sqrt{2} \to h/\mu = 0.81$, where the pairing gap jumps from $\Delta_0/\mu=1.162$ to zero~\cite{Gubbels}. Comparisons between different theoretical results obtained for the CC limit at unitarity can be found as, for example, in Refs.~\cite{Bernhard,Lang,JPB,Drut}.

It is worth mentioning that the exotic and elusive FFLO state (not considered here) is predicted to exist only in a very narrow window of $h$~(see, for instance, Ref.~\cite{Strinati} and references therein). Besides, the GMB correction shrinks the MF FFLO window, making it even more difficult to be unambiguously observed, at least in 3D~\cite{ADP}.

\begin{figure}
\centerline{\includegraphics[clip,width=3.7in]{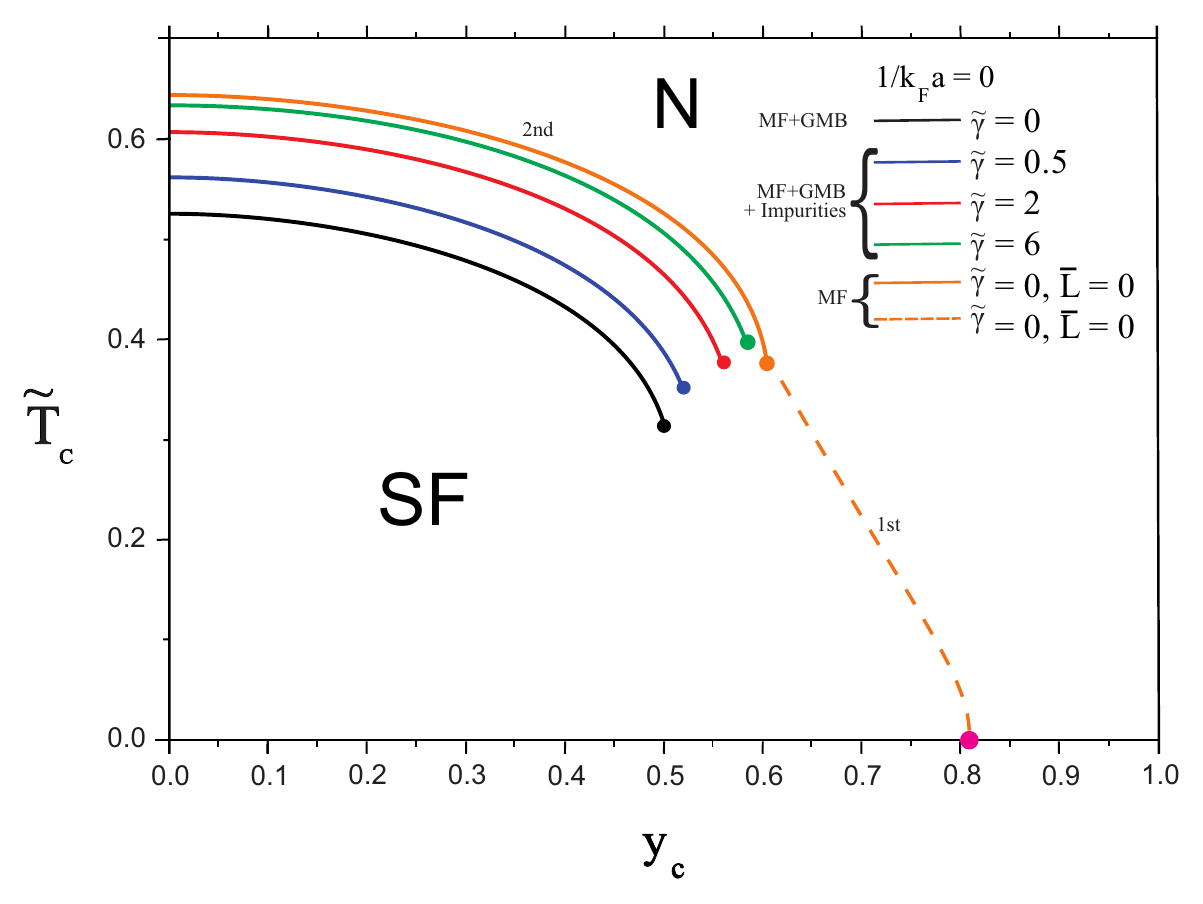}}
\caption{(Color online) Phase diagram of the imbalanced Fermi gas at unitarity $(k_Fa=\infty)$, for different values of the impurity parameter $\tilde \gamma$. All (solid line) curves describe the second order phase transition line between the superfluid (SF) and normal (N) phases, and stop at a tricritical point, below which the transition turns into a first-order. The top curve (orange) is the pure mean-field (with no GMB correction) standard phase boundary. The point $(0,0.657)$ at the $\tilde T_c$ axes is the critical temperature of the (balanced) BCS case. The tricritical point at the end of the MF second-order curve at $(y_c,\tilde T_c) \approx (0.63,0.37)$ is also shown. The MF first-order (dashed) line starts at the tricritical point and ends at the CC point $(y_{CC},0)=(0.81,0)$ (pink dot at the $y_c$ axes) where there is a (zero temperature) quantum phase transition from the SF to the N phase. The lower curve (black) is with the GMB correction in the clean limit ($\tilde \gamma=0$), and has the tricritical point at $(y_c,\tilde T_c) \approx (0.50,0.32)$. Notice that the GMB correction shrinks considerably the SF area, when compared to the area below the (top) MF curve. The curves above the lower curve are also corrected by the GMB correction, but were calculated in the presence of impurities. The second curve from bottom to top (blue) is for $\tilde \gamma=0.5$ and the next-one (red) is for $\tilde \gamma=2.0$. The fourth curve from bottom to top (green) is for $\tilde \gamma=6.0$. The effect of increasing the strength of the impurities is to cancel the GMB correction.}
\label{PD}
\end{figure}

\section{Effects of induced interactions corrections to the Chandrasekhar-Clogston limit of an imbalanced Fermi gas}
\label{CCL}

In order to have a quantitative description of the complete (first- and second-order) phase transitions, besides the second-order line and the tricritical point that we have investigated in the previous Section, the CC point also needs to be corrected accordingly. If the system is strongly interacting, the value of the CC field can also depend on the interactions in the normal state, an effect that must be accurately taken into account if we wish to study the N-SF phase transition. To be precise, one has to take into account self-energy effects in the normal state, mainly in the unitarity limit~\cite{Combescot,Punk1,Punk2,Baarsma}. This problem was first addressed in the context of the Fermi polaron, i.e., a small density of spin-down atoms interacting attractively with a non-interacting Fermi gas of spin-up atoms~\cite{Lobo}.

Thus, we are interested here in the Fermi polaron, the quasiparticle formed when few atoms interact with a fermionic environment of opposite spin, and we determine the quasiparticle properties of the Fermi polaron from the associated self-energy, through the use of the (non-selfconsistent~\cite{NSC}) T-matrix approximation, based on ladder diagrams~\cite{Combescot}.

Before we proceed, let us make some additional comments on the T-matrix approximation, which we have adopted to find the CC limit at unitarity. The T-matrix approach give results that coincide with a variational calculation~\cite{ChevyPRA,Combescot}, which in turn agree remarkably well with the QMC calculations~\cite{Giraud}. However, when there is a large mass asymmetry between the impurity (spin-down) atoms and that of the majority (spin-up) atoms, static impurities emerge in a Fermi sea of spin-$\uparrow$ fermions, signalizing the breakdown of the quasi-particle description~\cite{Bruun}. The quasi-particle as well as the Fermi-liquid picture of Fermi polarons through the many-body T-matrix theory in the strong-coupling limit also breaks down when the temperature of the system increases such that the Fermi surface broadens considerably, due to thermal fluctuations~\cite{Mulkerin,Mulkerin2}.

Then we investigate a small concentration of spin-down $\downarrow$ species of mass $m$ immersed in a spin-up $\uparrow$ Fermi sea of ultracold atoms of the same mass. The s-wave interaction potential between the very few spin-down species and the spin-up atoms in this unusual spin-imbalanced Fermi liquid is attractive. Specifically, we consider the case of a single spin-$\downarrow$ atom in a Fermi sea of spin-$\uparrow$ atoms. The Green's function of the minority species is given by means of the Dyson's equation,

\begin{eqnarray}
\label{FullProp}
 G_\downarrow^{-1}({\bf p},\omega) &=& G_{0 \downarrow}^{-1}({\bf p},\omega) - \Sigma({p},\omega)\\
 \nonumber
 &=& \omega - \epsilon_{p \downarrow} + \mu_\downarrow - \Sigma({p},\omega),
\end{eqnarray}
where $\epsilon_{p \downarrow} = p^2/2m$, and $\Sigma(q,i\omega_n)$ is the self-energy. The lowest energy for the spin-$\downarrow$ atom is expected to occur for $p = 0$, which gives rise to the expression $\mu_\downarrow = \Sigma(0,0)$ from the pole of Eq.~(\ref{FullProp}), as in~\cite{Combescot}.

The irreducible self-energy in the ladder approximation is given by~\cite{Fetter},

\begin{eqnarray}
\label{self-0}
\Sigma({p},i\omega_n) &=& T \sum_\nu \int \frac{d^3k}{(2\pi)^3}\\
\nonumber
&\times&  \Gamma({p},i\omega_\nu) G_{0\uparrow}({\bf p}-{\bf k},i\omega_\nu-i\omega_n), 
\end{eqnarray}
while the full vertex $\Gamma$ is given by the Bethe-Salpeter equation~\cite{Fetter}, which is a function of the total momentum $p$ and energy $\Omega$ of the incoming particles,

\begin{eqnarray}
\Gamma(p,\Omega) = \Gamma(p,\Omega)_0 + \Gamma(p,\Omega)_0 \Pi(p,\Omega) \Gamma(p,\Omega),
\label{vertex-1}
\end{eqnarray}
where $\Gamma(p,\Omega)_0 \equiv g$ is the bare coupling constant and $\Pi(p,\Omega)$ is the pair-propagator for two propagating atoms (or the two-particle self-energy). The above equation is easily solved for $\Gamma(p,\Omega)$ as

\begin{eqnarray}
\Gamma(p,\Omega)^{-1} = g^{-1} - \Pi(p,\Omega).
\label{vertex-2}
\end{eqnarray}
As before, $g$ can be traded by the physical s-wave scattering length $a$ characterizing the interaction between the single species and the background atoms, by using standard scattering theory. Thus, with Eq.~(\ref{ap11}), Eq.~(\ref{vertex-2}) turns out to be~\cite{Combescot,Punk1,Punk2},

\begin{eqnarray}
\Gamma(p,\Omega)^{-1} = \frac{m}{4 \pi a} - \left( \Pi(p,\Omega) + \int \frac{d^3 k}{(2 \pi)^3} \frac{1}{2 \epsilon_k} \right).
\label{vertex-4}
\end{eqnarray}

The pair-propagator is given by

\begin{eqnarray}
&&\Pi(p,\Omega) =  \\
\nonumber
&-& T \sum_n \int \frac{d^3 k}{(2\pi)^3}  G({\bf p} - {\bf k},-i\Omega-\omega_n) G({\bf k},\omega_n)=\\
\nonumber
&-& T \sum_n \int \frac{d^3 k}{(2\pi)^3}  \frac{G({\bf p} - {\bf k},-i\Omega-\omega_n) + G({\bf k},\omega_n)}{G({\bf p} - {\bf k},-i\Omega-\omega_n)^{-1} + G({\bf k},\omega_n)^{-1}}\\
\nonumber
&=&\int \frac{d^3 k}{(2\pi)^3} \left[ \frac{1-f_{\uparrow}(k) -f_{\downarrow}(p-k)}{\Omega+ i0^+ - \xi_{\uparrow k} - \xi_{\downarrow p- k}} \right],
\label{vertex-6}
\end{eqnarray}
where $f_{\sigma}(k)=[exp(\beta \xi_{\sigma k})+1]^{-1}$, with $\sigma = \uparrow, \downarrow$ and $\beta = 1/k_BT$ is the Fermi function, and $\xi_{\sigma k}=k^2/2m -\mu_{\sigma}$ is the single particle dispersion measured from the chemical potential. Next, we make the change of variables ${\bf k} \to  {\bf p} - {\bf k}$, and find

\begin{eqnarray}
\tilde \Pi(p,\Omega) =\int \frac{d^3 k}{(2\pi)^3} \left[ \frac{1-f_{\uparrow}(p-k) -f_{\downarrow}(k)}{\Omega+ i0^+ - \xi_{\uparrow p-k} - \xi_{\downarrow  k}} + \frac{m}{k^2}\right],
\label{vertex-7}
\end{eqnarray}
where we have defined $\tilde \Pi(p,\Omega)=\Pi(p,\Omega) + \int \frac{d^3 k}{(2 \pi)^3} \frac{1}{2 \epsilon_k}$. Taking the zero temperature limit and using the fact that $\mu_{\downarrow}$ is negative, we find in the limit where the ratio $\rho =  |\mu_{\downarrow}|/\mu_{\uparrow}$ is huge~\cite{Combescot},

\begin{eqnarray}
\label{Gamma}
\Gamma(0,0)^{-1} &=& \frac{m}{4 \pi a}  - \frac{m k_F}{4\pi^2} \pi \sqrt{\frac{\rho}{2}}\\
\nonumber
&=& \frac{m k_F}{4\pi} \left[ \frac{1}{k_F a}  -  \sqrt{\frac{\rho}{2}}  \right].
\end{eqnarray}

As we mentioned earlier, one of the problems of the T-matrix approximation is that this approximation neglects the particle-hole channel~\cite{Loktev}, which will be considered here. As we have seen before, by considering the induced interactions the inverse of the effective pairing interaction between atoms with different spins is modified as,
\begin{eqnarray}
\label{induce2}
\frac{1}{g} \to \frac{1}{U_\text{eff}} = \frac{m}{4 \pi a} + \chi_\text{ph} - \int \frac{d^3 k}{(2 \pi)^3} \frac{1}{2 \epsilon_k},
\end{eqnarray}
where $\chi_\text{ph}(p')$ is the polarization function, and we have made use of Eq.~(\ref{ap11}). As we have seen earlier, the generalized static polarization function is given by
$\chi_\text{ph}(x,y,\tilde \gamma) = - N(0) L(x,y,\tilde \gamma)$, where, as we have defined before, $x\equiv \frac{q}{2 k_\text{F}}$, $y\equiv \frac{h}{E_F}$, $\tilde \gamma = \frac{\gamma}{E_F}$, $N(0)=mk_F/(2\pi^2)$ is the density of states at the Fermi level, and $L(x,y,\tilde \gamma)$ is the $y$ and $\tilde \gamma$-dependent Lindhard screening function.

Taking into account the induced interactions, Eq.~(\ref{Gamma}) is modifed to

\begin{eqnarray}
\label{vertex-11-2}
\Gamma(0,0)^{-1} &=& \frac{m}{4 \pi a} + \tilde \chi_\text{ph}  - \frac{m k_F}{4\pi^2} \pi \sqrt{\frac{\rho}{2}}\\
\nonumber
&=& \frac{m k_F}{4\pi} \left[ \frac{1}{k_F a} - \frac{2 \tilde L}{\pi}  -  \sqrt{\frac{\rho}{2}}  \right],
\end{eqnarray}
where $\tilde L$ comes from the angle average of the polarization function $\tilde{\chi}_\text{ph}(x,y) \equiv \langle {\chi}_\text{ph}(x,y)\rangle = \langle - N(0) L(x,y) \rangle= - N(0) \tilde L(y)$. For $h=0$, the well known result is $\tilde{\chi}_\text{ph} = - N(0) (1+2\ln 2)/3 = - N(0) \ln(4e)^{1/3} \equiv - N(0) \tilde L_0$, where $\tilde L_0 \equiv \tilde L(0)=\ln(4e)^{1/3}$.

The spin-$\uparrow$ atom Green's function in Eq.~(\ref{self-0}), is the bare one, $G_{0\uparrow}(k,\omega)=[\omega - \epsilon_{k \uparrow} +\mu_{\uparrow}]^{-1}$, where $\mu_{\uparrow}=k_F^2/2m_{\uparrow}$. $G_{0\uparrow}(k,\omega)$ has a single pole at $\omega = \epsilon_{k \uparrow} - \mu_{\uparrow}$. After the contour $C$ integration, the self-energy in Eq.~(\ref{self-0}) reads~\cite{Combescot}

\begin{eqnarray}
\Sigma({p},\omega) &=&  \int \frac{d^3k}{(2\pi)^3} \theta (\mu_{\uparrow} - \epsilon_{k-p \uparrow})  \Gamma(k,\omega + \epsilon_{k-p \uparrow} - \mu_{\uparrow} )
\nonumber
\\
&=&  \int_0^{k_F} \frac{dk~k^2}{2\pi^2} \langle \Gamma(k+p,\omega + \epsilon_{k \uparrow} - \mu_{\uparrow} ) \rangle,
\label{Self-0}
\end{eqnarray}
where the bracket $\langle~ \rangle$ in the equation above is for the angular average over the direction of $k$. Using $\Gamma$ for the case obtained before where $\rho$ is large, the above equation turns out to be,

\begin{eqnarray}
\Sigma({p},\omega) &=&  \Gamma(0,0) \int_0^{k_F} \frac{dk~k^2}{2\pi^2} = \frac{k_F^3}{6\pi^2} \Gamma(0,0),
\label{Self-1}
\end{eqnarray}
where $\Gamma(0,0)$ is given by Eq.~(\ref{vertex-11-2}). Then, Eq.~(\ref{FullProp}) yields

\begin{eqnarray}
\mu_{\downarrow} &=& \frac{k_F^3}{6\pi^2} \Gamma(0,0).
\label{Self-2}
\end{eqnarray}
Plugging $\Gamma(0,0)$ from Eq.~(\ref{vertex-11-2}) into Eq.~(\ref{Self-2}), we find

\begin{eqnarray}
\frac{1}{k_F a} = \sqrt {\frac{\rho}{2}} - \frac{2}{3 \pi} \frac{2}{\rho} + \frac{2 \tilde L}{ \pi}.
\label{Self-4}
\end{eqnarray}
At unitarity, $\rho$ is found as the real solution of the equation

\begin{eqnarray}
\rho^3 - 2 \left( \frac{2 \tilde L}{\pi}\right)^2 \rho^2 + \frac{32 \tilde L}{3 \pi^2}\rho - 2  \left( \frac{4}{3 \pi} \right)^2  = 0.
\label{Self-5}
\end{eqnarray}
\\
{\bf Clean Regime}\\
\\
We calculate now the CC field taking into account the GMB correction in the clean limit, characterized by $\tilde \gamma=0$. For $\tilde L=0$ i.e., without the GMB correction, $\rho = 2^{1/3}  \left( \frac{4}{3 \pi} \right)^{2/3} \approx 0.71$, in complete agreement with \cite{Combescot}. Considering the screening of the interaction, $\rho$ turns out to be dependent on $\tilde L $. As we have seen previously (see Fig.~\ref{L}), this function varies from the maximum value obtained in the balanced situation $\tilde L (y=0) = 0.795$ to its lower value, which is for the fully polarized case $\tilde L (y=1,\tilde \gamma=0) = 0.69$~\cite{ADP}. Considering the ``polaronic limit'' as that of $\tilde L (y=1,\tilde \gamma=0)$, we get $\rho \approx 0.46$.

Theoretical and experimental investigations have found that at unitarity the chemical potential and mass of the single atom are shifted by $\mu_\downarrow \to \mu_p = A \mu_\uparrow$ and $m \to m^*$, where $\mu_p$ is the polaron chemical potential and $m^*$ is its effective mass. In the present work, $A$ is ratio we have just found, $A = - 0.46$. It can be shown that the chemical potential of the minority species is related to that of the polaron as~\cite{Mora},

\begin{eqnarray}
\mu_\downarrow = \mu_p + E_{F \downarrow},
\label{Self-6}
\end{eqnarray}
where the Fermi energy $E_{F \downarrow}=k_{F \downarrow}^2/2m^*$ of the spin-down atom is vanishingly small.

As in Refs.~\cite{Navon,Mora} we write the equation of the pressure of an intermediate normal imbalanced (N-I) state (i.e., not a polarized normal state), as a sum of the (individual) pressures of an ideal gas of majority atoms with chemical potential $\mu_{\uparrow}$ and an ideal gas of polarons with chemical potential given at unitarity by $\mu_p = A\mu_{\uparrow}$, with $A=-0.46$,

\begin{widetext}
\begin{eqnarray}
\label{PNM}
P^{\rm N-I}(\mu_{\downarrow\uparrow}) &=& \frac{1}{15 \pi^2} (2m)^{3/2} \mu_{\uparrow}^{5/2} + \frac{1}{15 \pi^2} (2m^*)^{3/2} (\mu_{\downarrow}-\mu_{p})^{5/2},
\\
\nonumber
&=& \frac{1}{15 \pi^2} (2m)^{3/2} \mu_{\uparrow}^{5/2} \left[1 + \left(\frac{m^*}{m} \right)^{3/2} (\eta + |A|)^{5/2} \right],
\end{eqnarray}
\end{widetext}
where $m^*$ is the effective polaron mass, and $\eta \equiv \mu_{\downarrow}/ \mu_{\uparrow}$.

The ``Universality Hypothesis'' at the unitary limit, which relays on a universal thermodynamics~\cite{Ho}, states that the pressure in the superfluid phase (SP) can be written as~\cite{ChevyPRL,ChevyPRA}

\begin{eqnarray}
\label{PS}
P^{\rm S}(\mu)&=& \frac{1}{15 \pi^2} \left(\frac{m}{\xi}\right)^{3/2}  \left(\mu_{\uparrow} + \mu_{\downarrow} \right)^{5/2} 
\\
\nonumber
&=& \frac{1}{15 \pi^2} \left(\frac{m}{\xi}\right)^{3/2} \left(2\mu \right)^{5/2}
\\
\nonumber
&=& \frac{1}{15 \pi^2} \left(\frac{2m}{2\xi}\right)^{3/2} \mu_{\uparrow}^{5/2} \left(1+\eta \right)^{5/2},
\end{eqnarray}
where $\xi \sim 0.42(1)$, is the universal Bertsch parameter, obtained both theoretically (quantum Monte Carlo (QMC)) and experimentally~\cite{Boronat,Sanjay}.

From equilibrium conditions, $\mu=(\mu_{\uparrow} + \mu_{\downarrow})/2$ and $P^{\rm S}(\mu)= P^{\rm N-M}(\mu_{\downarrow\uparrow})$, between the N-I and S phases (and assuming a disregardable surface tension in the boundary between the N-I and S regions~\cite{ST1,ST2,ST3,ST4}) we obtain

\begin{eqnarray}
\label{PS3}
1 + \left(\frac{m^*}{m} \right)^{3/2} (\eta + |A|)^{5/2}=\frac{1}{\left(2\xi \right)^{3/2}} \left(1+\eta \right)^{5/2}.
\end{eqnarray}

Different calculations and methods give slight disagreements on the value of the effective mass~\cite{Mora}. Systematic diagrammatic expansion for example, yields $m^*=1.2m$~\cite{Giraud} which is in agreement with the value found in the following experiments $m^*=1.20(2)m$~\cite{FP1,FP2,Navon}. A recent experiment~\cite{FP7} has found $m^*=1.25(5)m$, agreeing, among other theoretical results, with results obtained from diagrammatic Monte Carlo calculations~\cite{Svistunov}. We solve this equation using the experimental value, $m^*=1.25m$, and find

\begin{eqnarray}
\label{Equilibrium3}
\eta_{c} \sim -0.046.
\end{eqnarray}

\begin{figure}
\centerline{\includegraphics[clip,width=3.2in]{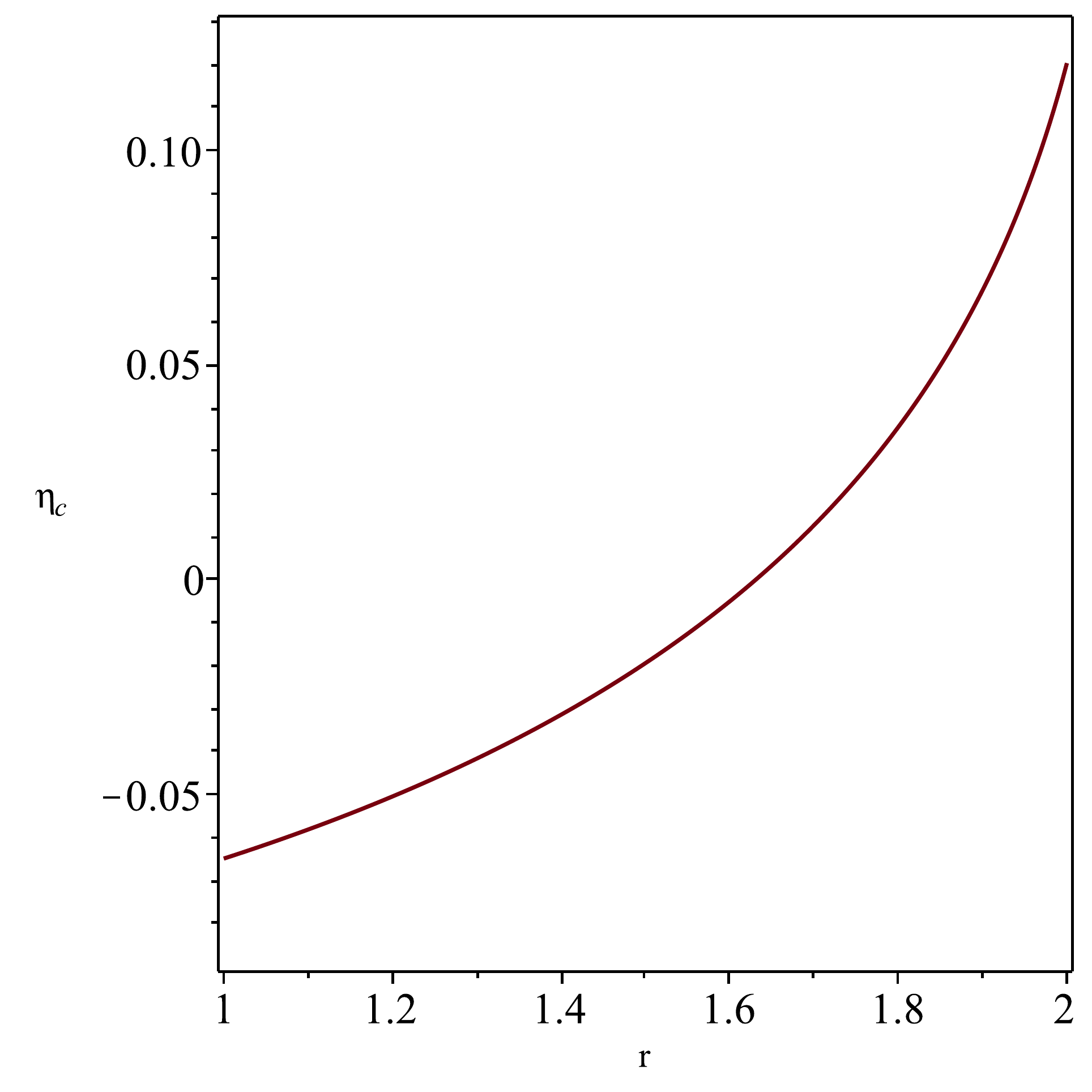}}
\caption{(Color online) Behavior of $\eta_c$ as a function of the ratio $r = m^*/m$ of the polaron effective mass by the bare spin-down atom mass in the interval $1 \leq r \leq 2$. For $r=1$, $\eta_c=-0.065$; for $r=1.63$, $\eta_c=0$ and finally, for $r=2$, $\eta_c=0.12$.}
\label{etaVSr}
\end{figure}

Although we use the experimental value $m^*=1.25m$ in our calculations, for completeness in Fig.~\ref{etaVSr} we show how the critical $\eta_c$ depends on the mass ratio $r \equiv m^*/m$. Our choice to investigate the behavior of $\eta_c$ with maximum $r$ up to $2$ is due to the fact that an effective mass larger than $2m$ has been found when the nature of the quasi-particle changes from a Fermi polaron (fermionic quasi-particle) to a molecule (bosonic quasi-particle)~\cite{Svistunov}.

The ratio $\eta$ of the chemical potentials is given by

\begin{eqnarray}
\label{ratioeta1}
\eta = \frac{\mu_{\downarrow}}{\mu_{\uparrow}}=\frac{1-h/\mu}{1+h/\mu}.
\end{eqnarray}
From Eq.~(\ref{ratioeta1}) one easily finds

\begin{eqnarray}
\label{saturation1}
\frac{h}{\mu} = \frac{1-\eta}{1+\eta}.
\end{eqnarray}

With the equation above we obtain $(h/\mu)_{CC}=y_{CC}$ as a function of $\eta_{c}$ given by Eq.~(\ref{Equilibrium3}),

\begin{eqnarray}
\label{PS4}
y_{CC} = 1.09,
\end{eqnarray}
which is the CC field obtained with the T-matrix approximation at the unitary limit, taking into account the GMB screening of the interactions without impurities.

We point out that although the current experimental value of the polaron effective mass $m^*$ is found to be close to the bare minority species mass $m$, the value of $y_{CC}$ depends significantly on the ratio $r=m^*/m$. For $r= 1$ for instance, we would obtain $y_{CC} = 1.14$, for $r=1.5$ we have $y_{CC} = 1.04$ and for $r=2$ we find $y_{CC} = 0.78$. This can be summarized as: for $1 \leq r \leq 1.63$ we have $\eta_c \leq 0$ with $y_{CC}(r=1)=1.14$ and $y_{CC} (r=1.63)= 1$; and for $1.63 \leq r \leq 2$, $\eta_c \geq 0$ with $y_{CC}(r=2)=0.78$.\\
\\
{\bf Impurity Regime}\\
\\
Let us now evaluate the CC field taking into account the GMB correction in the presence of impurities, identified by a finite $\tilde \gamma$. Still in the polaronic limit ($y \to 1$), and taking into consideration here the following values of the impurity parameter: $\gamma=0.5$, $\gamma=0.2$, and $\gamma=4.0$. 

For $\gamma=0.5$, we find $\tilde L (y=1,\tilde \gamma=0.5)=0.61$, which yields $\rho = 0.48$. From Eq.~(\ref{PS3}) we find

\begin{eqnarray}
\label{Equilibrium-1}
\eta_{c} \sim -0.03677,
\end{eqnarray}
which gives,

\begin{eqnarray}
\label{PS5}
y_{CC} = 1.076.
\end{eqnarray}

For $\gamma=2.0$ we obtain $\tilde L (y=1,\tilde \gamma=2.0)=0.3696$, resulting in $\rho = 0.55$, that, with Eq.~(\ref{PS3}) we obtain

\begin{eqnarray}
\label{Equilibrium-2}
\eta_{c} \sim 0.00799,
\end{eqnarray}
which implies in

\begin{eqnarray}
\label{PS6}
y_{CC} = 0.98.
\end{eqnarray}

And finally, for $\gamma=4.0$ we obtain $\tilde L (y=1,\tilde \gamma=4.0)=0.2045$, resulting in $\rho = 0.6184$, that, again with Eq.~(\ref{PS3}) we obtain

\begin{eqnarray}
\label{Equilibrium-3}
\eta_{c} \sim 0.0769,
\end{eqnarray}
which implies in

\begin{eqnarray}
\label{PS7}
y_{CC} = 0.857.
\end{eqnarray}

All values calculated for $\rho$ are in relatively good agreement with the available QMC result $\rho = 0.58 \pm 0.01$~\cite{ChevyPRA,Lobo}. Among them the closest one was that found for $\gamma=2.0$, which gave $\rho = 0.55$.

Our values of $y_{CC}$ for both clean and with impurities are within the results found by means of several theoretical approaches as, for instance, Lobo et al.~\cite{Lobo}, who found $(h/\mu)_c=0.96$ from QMC calculations; Boettcher et al.~\cite{Boettcher}, who obtained $(h/\mu)_c=0.83$ by means of the functional renormalization group (FRG). With the FRG they were able to go beyond the mean-field approximation by including bosonic fluctuations on the many-body state; Land et al.~\cite{Lang}, who found $(h/\mu)_c=1.09$ within a Luttinger-Ward formalism; and $(h/\mu)_c=0.88$ in the work by Caldas~\cite{JPB}, obtained by means of thermodynamic equilibrium between possible phases at the unitary limit. We would like to remark in passing, for reference only, that the experimental value obtained for the zero temperature CC limit of a (harmonically trapped) spin-polarized Fermi gas of ${}^{6} \rm Li$ atoms at unitarity is $(h/\mu)_c \simeq 0.95$~\cite{Exp1,Shin-Ketterle}.

\section{Current and some Possible Experiments with Impurities in Cold Atoms}
\label{PE}

Besides the case of considering a spin-$\downarrow$ atom in a sea of spin-$\uparrow$ atoms of the same species discussed in Section~\ref{CCL} as an ``impurity'' (forming a Fermi polaron), which has been observed in~\cite{FP1,FP2,FP3,FP4,FP5,FP6,FP7}, other very interesting pictures of impurities in cold atoms are in progress.

Regarding atoms of different elements, it is possible to introduce impurities in cold atoms by doping the superfluid with a different species or spin-state. However, the effects of this doping are different from what happens in solid state materials, where the impurity is often used to control the electron density by making the impurity a donor or an acceptor. In cold atoms the impurities would possibly play a different role, such as mediating interactions between the background atoms. In this case, the impurity would either help pairing or hinder it~\cite{Randy}.

Another possibility was reported very recently with atoms of different species, where Bose polarons were created near quantum criticality by immersing atomic impurities (potassium) of both spin states in a Bose-Einstein condensate (BEC) of sodium atoms with near-resonant interactions, at zero and non-zero temperature~\cite{BP}.

\section{Conclusion}
\label{Conc}
 
In summary, we have investigated in homogeneous 3D Fermi systems the GMB corrections beyond MF to the transition temperature $T_c$, in the case of a balanced Fermi gas, and to the tricritical point in the case of an imbalanced gas of fermionic atoms at unitarity. In the calculations we considered the presence of (nonmagnetic) impurities or defects, in terms of a finite lifetime $\tau = 1/\gamma$ of the particle-hole fluctuations of the medium.

In the case of balanced Fermi gas, we find that the GMB correction results in the well known reduction in the MF $T_c$ by a factor of $\approx 2.2$. However, the presence of impurities is deleterious for the GMB corrections, such that by increasing the amount of impurity, these corrections are completely nullified, and one is left with the MF result.

For imbalanced Fermi gases, we find that the GMB correction is maximum in the clean limit, and pulls down the entire second-order transition line and the tricritical point. As happens for the balanced case, impurities lead to cancel the GMB correction and suspend the second-order transition curve, and consequently the tricritical point $(y_{c},\tilde T_{c})$. A high concentration of impurities tends to push the second-order curve back to the mean-field result.

We also found the first-order zero temperature CC quantum phase transition. In order to find a reliable value of the critical field $y_{CC}$ for the CC-transition of the unitary imbalanced Fermi gas, we employed a T-matrix approach, based on ladder diagrams, and also considered the GMB screening of the interactions in the pure and impurity regimes. We obtained very good agreement with known QMC and previous theoretical results.

A possible and natural extension of our results could be to consider pairing and screening in the GMB correction~\cite{Pisani} to the critical temperature throughout the whole BCS-BEC crossover of an imbalanced Fermi gas.

\appendix

\section{Calculation of the pair susceptibility $\chi_\text{ph}$ in the presence of impurities}
\label{ApA}

After performing the integration in the angles in Eq.~(\ref{chi0}), we obtain
\begin{widetext}
\begin{eqnarray}
\label{chiph-5}
Re \chi_\text{ph}(q,h,\gamma)
=& - &\frac{2m}{(2 \pi)^2} \int \frac{dk~k^2}{4qk}  \\
\nonumber
&\times& \left[  f_{k}^{\downarrow} \ln \left( \frac{(q^2 -4mh -2m \omega_0+ 2kq )^2  + (2m\gamma)^2 }{(q^2 -4mh - 2m \omega_0 -2kq)^2 + (2m\gamma)^2} \right)+   f_{k}^{\uparrow} \ln \left( \frac{(q^2 + 4mh + 2m \omega_0+ 2kq )^2  + (2m\gamma)^2 }{(q^2 + 4mh + 2m \omega_0 -2kq)^2 + (2m\gamma)^2} \right)   \right].
\end{eqnarray}
where $q \equiv |\vec q|$. Equation~(\ref{chiph-5}) is usually calculated in the zero temperature limit. Then, at $T\to0$ we have $ f_{k}^{\downarrow,\uparrow} \to  \Theta(k_\text{F}^{\downarrow,\uparrow}-k)$, where $\Theta(x)$ is the Heaviside step function, such that the correction to the coupling $g$ from the static induced interaction is a (temperature independent) constant,
\begin{eqnarray}
\label{chiph-6}
Re \chi_\text{ph}( q,h,\gamma) &\equiv& Re \chi_\text{ph}(q,h,\gamma)^{\downarrow} + Re \chi_\text{ph}(q,h,\gamma)^{\uparrow} =\\
\nonumber
&- &\frac{m}{2(2 \pi)^2 q} \left[ \int_0^{k_F^{\downarrow}} dk~k    \ln \left( \frac{(q^2 -4mh+2kq)^2 + (2m\gamma)^2 }{(q^2 -4mh-2kq)^2+ (2m\gamma)^2} \right)+   \int_0^{k_F^{\uparrow}} dk~k  \ln \left( \frac{(q^2 +4mh+2kq)^2+ (2m\gamma)^2 }{(q^2 +4mh-2kq)^2+ (2m\gamma)^2} \right)   \right].
\end{eqnarray}
Equation \ref{chiph-6} clearly shows that for a spin imbalanced Fermi gas the static polarization function is given by separated contributions from spin-down and spin-up-like susceptibilities. The above equation may be rearranged as

\begin{eqnarray}
\label{chiph-lf-8}
Re \chi_\text{ph}(q,h,\gamma) &\equiv& Re \chi_\text{ph}(q,h,\gamma)^{\downarrow} + Re \chi_\text{ph}(q,h,\gamma)^{\uparrow} \\
\nonumber
&=&- \frac{m}{2(2 \pi)^2 q} \left[ \int_0^{k_F^{\downarrow}} dk~k    \ln \left( \frac{(A+k)^2 + C^2}{(A-k)^2+ C^2} \right)
+   \int_0^{k_F^{\uparrow}} dk~k  \ln \left( \frac{(B+k)^2+ C^2 }{(B-k)^2+ C^2} \right)   \right],
\end{eqnarray}
where $A \equiv \frac{q^2 -4mh}{2q}$, $B \equiv \frac{q^2 +4mh}{2q}$ and $C \equiv \frac{m\gamma}{q}$, which after integration in $k$ yields

\begin{eqnarray}
\label{chiph-lf-9}
Re \chi_\text{ph}(q,h,\gamma) &\equiv& Re \chi_\text{ph}(q,h,\gamma)^{\downarrow} + Re \chi_\text{ph}(q,h,\gamma)^{\uparrow} = \\
\nonumber
&-& \frac{m}{2(2 \pi)^2 q} \Big\{  \frac{1}{2} \left( {k_F^{\downarrow}}^2 - (A^2-C^2) \right)  \ln \left| \frac{(A+ k_F^{\downarrow})^2 + C^2}{(A-k_F^{\downarrow})^2+ C^2} \right| + 2 A k_F^{\downarrow} \\
\nonumber
&-& 2AC \left[ \arctan \left( \frac{k_F^{\downarrow}-A}{C} \right) + \arctan \left( \frac{k_F^{\downarrow}+A}{C} \right) \right] \big] \\
\nonumber
&-& \frac{m}{2(2 \pi)^2 q} \big[  \frac{1}{2} \left( {k_F^{\uparrow}}^2 - (B^2-C^2) \right)  \ln \left| \frac{(B+ k_F^{\uparrow})^2 + C^2}{(B-k_F^{\uparrow})^2+ C^2} \right| +2 B k_F^{\uparrow} \\
\nonumber
&-& 2BC \left[ \arctan \left( \frac{k_F^{\uparrow}-B}{C} \right) + \arctan \left( \frac{k_F^{\uparrow}+B}{C} \right) \right] \Big\}. \\
\end{eqnarray}

Plugging $A, B$ and $C$ into $Re \chi_\text{ph}(q,h,\gamma)$ gives

\begin{eqnarray}
\label{chiph-7}
\chi_\text{ph}^{\downarrow}(q,h,\gamma) =&-& \frac{m}{ 8 \pi^2 q} \Big\{  \frac{1}{2} \left[ {k_\text{F}^{\downarrow}}^2-\left( \frac{q^2-4mh}{2q} \right)^2 + \left(\frac{m\gamma}{q} \right)^2 \right ]\ln \left| \frac{(q^2 -4mh+2q k_\text{F}^{\downarrow})^2 + \left(2m\gamma \right)^2 }{(q^2 -4mh-2q k_\text{F}^{\downarrow})^2 + \left(2m\gamma \right)^2} \right| \\
\nonumber
&+& k_\text{F}^{\downarrow}\left( \frac{q^2-4mh}{q} \right) \\
\nonumber
&-& 2 \left[ \left( \frac{q^2 - 4mh}{2q} \right) \frac{m\gamma}{q} \right] \left[ \arctan \left( \frac{2q k_F^{\downarrow}-q^2 + 4mh}{2 m \gamma} \right) + \arctan \left( \frac{2q k_F^{\downarrow}+q^2 - 4mh}{2m\gamma} \right) \right] \Big\}, \\
\label{chiph-8}
\chi_\text{ph}^{\uparrow}(q,h,\gamma) =&-& \frac{m}{ 8 \pi^2 q} \Big\{  \frac{1}{2}  \left[ {k_\text{F}^{\uparrow}}^2-\left( \frac{q^2+4mh}{2q} \right)^2 + \left(\frac{m\gamma}{q} \right)^2 \right ] \ln \left| \frac{(q^2 +4mh+2q k_\text{F}^{\uparrow})^2 + \left(2m\gamma \right)^2 }{(q^2 +4mh-2q k_\text{F}^{\uparrow})^2 + \left(2 m\gamma \right)^2} \right| \\
\nonumber
&+& k_\text{F}^{\uparrow}\left( \frac{q^2+4mh}{q} \right)  \\
\nonumber
&-& 2 \left[ \left( \frac{q^2 + 4mh}{2q} \right) \frac{m\gamma}{q} \right] \left[ \arctan \left( \frac{2q k_F^{\uparrow}- q^2 - 4mh}{2m\gamma} \right) + \arctan \left( \frac{2q k_F^{\uparrow}+q^2 +4mh}{2m\gamma} \right) \right] \Big\}.
\end{eqnarray}
The above equations can be written in a more appropriate form, according to the definition $\tilde{\chi}_\text{ph}^\sigma(x,y,\tilde \gamma)\equiv \chi_\text{ph}^\sigma(q,h,\gamma)$, with
\begin{eqnarray}
\label{chiph-full-1}
&& \tilde{\chi}_\text{ph}^{\downarrow}(x,y,\tilde \gamma) \equiv - N(0) L^{\downarrow}(x,y,\tilde \gamma) = - \frac{N(0)}{4} \Big\{  \sqrt{1-y} \left(1 - \frac{y}{2x^2} \right) \\
\nonumber
&-&  \frac{1}{2x} \left[ 1 - y - x^2 \left( 1- \frac{y}{2x^2} \right)^2 + \left( \frac{\tilde \gamma }{4 x}\right)^2 \right] \frac{1}{2} \ln \left| \frac{(\sqrt{1-y} + \frac{y}{2x} - x)^2 + \left( \frac{\tilde \gamma }{4 x}\right)^2}{(\sqrt{1-y} - \frac{y}{2x} + x)^2 + \left( \frac{\tilde \gamma }{4 x}\right)^2} \right|  \\
\nonumber
&-&  \left[ \left( 1- \frac{y}{2x^2} \right) \left( \frac{\tilde \gamma }{4 x}\right) \right] \left[ \arctan \left( \frac{4 x \sqrt{1-y}-4x^2 +2y}{ \tilde \gamma} \right) + \arctan \left( \frac{4 x \sqrt{1-y}+4x^2 -2y}{ \tilde \gamma}  \right) \right] \Big\}, \\
\label{chiph-full-2}
&& \tilde{\chi}_\text{ph}^{\uparrow}(x,y,\tilde \gamma) \equiv - N(0) L^{\uparrow}(x,y,\tilde \gamma) = - \frac{N(0)}{4} \Big\{  \sqrt{1+y} \left(1 + \frac{y}{2x^2} \right) \\
\nonumber
&-&  \frac{1}{2x} \left[ 1 + y - x^2 \left( 1+ \frac{y}{2x^2} \right)^2 + \left( \frac{\tilde \gamma }{4 x}\right)^2 \right] \frac{1}{2} \ln \left| \frac{(\sqrt{1+y} - \frac{y}{2x} - x)^2 + \left( \frac{\tilde \gamma }{4 x}\right)^2}{(\sqrt{1+y} + \frac{y}{2x} + x)^2 + \left( \frac{\tilde \gamma }{4 x}\right)^2} \right|  \\
\nonumber
&-&  \left[ \left( 1+ \frac{y}{2x^2} \right) \left( \frac{\tilde \gamma }{4 x}\right) \right] \left[ \arctan \left( \frac{4 x \sqrt{1+y}-4x^2 -2y}{ \tilde \gamma}  \right) + \arctan \left( \frac{4 x \sqrt{1+y}+4x^2 +2y}{\tilde \gamma}  \right) \right] \Big\},
\end{eqnarray}
where $x\equiv \frac{q}{2 k_\text{F}}=\sqrt{2(1+\cos \phi)}/2$, $y\equiv \frac{h}{E_F}$, and $\tilde \gamma \equiv \frac{\gamma}{E_F}$, $N(0)=mk_F/(2\pi^2)$ is the density of states at the Fermi level per spin component, $k_F=(3\pi^2 n)^{1/3}$ is the Fermi wave vector related to the particle density $n$, and $E_F=k_F^2/(2m)$ is the Fermi energy.

The simplified notation in $\tilde{\chi}_\text{ph}^\sigma(x,y,\tilde \gamma)$ permits us to express the polarization function of an imbalanced Fermi gas in the presence of non-magnetic impurities in a compact form as
\begin{eqnarray}
\label{generalizedLFinal}
\tilde{\chi}_\text{ph}(x,y,\tilde \gamma) = - N(0) L(x,y,\tilde \gamma),
\end{eqnarray}
where $L(x,y,\tilde \gamma) \equiv L^{\downarrow}(x,y,\tilde \gamma) + L^{\uparrow}(x,y,\tilde \gamma)$ is the generalized Lindhard function.

Taking the limit $\tilde \gamma \to 0$ in Eqs.~\ref{chiph-full-1} and~\ref{chiph-full-2} it is straightforward to obtain,

\begin{eqnarray}
\label{chiph7-2}
\tilde{\chi}_\text{ph}^{\downarrow}(x,y) &=& -\frac{N(0)}{4} \left\{\sqrt{1-y}\left(1 - \frac{y}{2x^2} \right)  -  \frac{1}{2x} \left[ 1 - y - x^2 \left( 1- \frac{y}{2x^2} \right)^2 \right ]  \ln \left| \frac{\sqrt{1-y} + \frac{y}{2x} - x }{\sqrt{1-y} - \frac{y}{2x} + x} \right|   \right\} \\
\nonumber
&\equiv& - N(0) L^{\downarrow}(x,y),
\end{eqnarray}
and
\begin{eqnarray}
\label{chiph8-2}
\tilde{\chi}_\text{ph}^{\uparrow}(x,y) &=& -\frac{N(0)}{4} \left\{  \sqrt{1+y}\left(1 + \frac{y}{2x^2} \right)    -  \frac{1}{2x} \left[ 1 + y - x^2 \left( 1+ \frac{y}{2x^2} \right)^2 \right ] \ln \left| \frac{\sqrt{1+ y} - \frac{y}{2x} - x }{\sqrt{1+ y} + \frac{y}{2x} + x} \right|  \right\} \\
\nonumber
&\equiv& - N(0) L^{\uparrow}(x,y),
\end{eqnarray}
\end{widetext}
which was obtained in the clean limit in Ref.~\cite{ADP}.\\
\newline

\section{Non-dimensional equality for the obtention of the first-order line and the thermal gap-equation}
\label{ApB}

The first-order line in the phase diagram in Fig.~\ref{PD} is obtained as the solution of the non-dimensional equality

\begin{widetext}
\begin{eqnarray}
\label{firstorder-1}
&-&\left[    \frac{\pi}{4 k_F a} -\frac{ \bar L(y,\tilde \gamma)}{2} -\int_0^Z \frac{dz}{2} \right] \tilde\Delta_0^2+ \int_0^Z z^2 dz \Big[ z^2-\tilde \mu - \sqrt{(z^2-\tilde \mu)^2 + \tilde \Delta_0^2 }  \\
\nonumber
&-& \tilde T \ln \left( e^{-\frac{1}{\tilde T}(\sqrt{(z^2-\tilde \mu)^2 + \tilde \Delta_0^2 }+y)} + 1\right) 
-\tilde T \ln \left( e^{-\frac{1}{\tilde T}(\sqrt{(z^2-\tilde \mu)^2 + \tilde \Delta_0^2 }-y)} + 1\right) \Big]\\
\nonumber
&=& \int_0^Z z^2 dz \left\{ z^2-\tilde \mu - |z^2- \tilde \mu| - \tilde T \ln \left( e^{-\frac{1}{\tilde T}(|z^2- \tilde \mu|+y)} + 1\right) 
-\tilde T \ln \left( e^{-\frac{1}{\tilde T}(|z^2- \tilde \mu|-y)} + 1\right) \right\},
\end{eqnarray}
\end{widetext}
together with the thermal gap equation

\begin{widetext}
\begin{eqnarray}
\label{firstorder-2}
 &&  \frac{\pi}{2 k_F a} - \bar L(y,\tilde \gamma) \\
   \nonumber
&+& \int_0^Z z^2 dz  \left[  \frac{1}{\sqrt{(z^2-\tilde \mu)^2 + \tilde \Delta_0^2 } } \left[1- \frac{1}{e^{\frac{1}{\tilde T}(\sqrt{(z^2-\tilde \mu)^2 + \tilde \Delta_0^2 }+y)}+1}  - \frac{1}{e^{\frac{1}{\tilde T}(\sqrt{(z^2-\tilde \mu)^2 + \tilde \Delta_0^2 }-y)}+1 } \right] -\frac{1}{z^2} \right]=0,
\end{eqnarray}
\end{widetext}
where $\bar L(y,\tilde \gamma) =  \frac{1}{2} \int_{-1}^{1} \mathrm{d} \cos\theta ~ L(x,y,\tilde \gamma)$, with $L(x,y,\tilde \gamma)$ given by Eq.~(\ref{generalizedLFinal}), $\tilde \mu \equiv \mu/E_F$, $Z \equiv \Lambda/k_F$, and $\tilde\Delta_0 = \Delta_0/E_F$.

\textbf{Acknowledgements} \par 

We would like to thank H. T. C. Stoof for many helpful and enlightening discussions, and for a critical reading on the preliminary version of the present manuscript. We are also grateful to Q. Chen, R. Combescot, W. Zwerger, M. Punk, B. Frank, R. Hulet, M. Zwierlein and B. Mulkerin for useful and fruitful conversations. H.~C. wish to thank CNPq and FAPEMIG for partial financial support.


\begin{thebibliography}{41}
\bibitem{AndersonTheorem} Anderson's theorem states that superconductivity in a conventional superconductor is robust with respect to non-magnetic weak disorder in the host material. Conventional superconductors are superconducting materials described by the BCS mean-field theory i.e., all the ones in which pairing is phonon mediated.
\bibitem{note} The divergence occurs at the resonance (the unitary limit). The unitary limit also corresponds to the threshold interaction strength at which a bound state appears in vacuum.
\bibitem{gen} In fact in Ref.~\cite{Pisani} it was used a partially self-consistent T-matrix approximation, on top of which the GMB corrections have been included. Five variants of the T-matrix approximation for an attractive dilute Fermi gas in its normal phase above the superfluid critical temperature $T_c$ have been systematically investigated, and both qualitative and quantitative differences among the results from the several approaches have been found~\cite{Pini}.
\bibitem{NSC} The designation ``non-selfconsistent'' means that the T-matrix approximation is constructed (i.e. all internal propagator lines in the diagrams) with ``bare'' (non-interacting) single-particle Green's functions~\cite{Punk2,Perali}.
\end{thebibliography}

%

\end{document}